\documentclass[journal]{IEEEtran}
\usepackage{booktabs}
\usepackage{multirow}
\usepackage{xcolor}
\usepackage{microtype}
\usepackage{multirow}
\usepackage{booktabs}
\usepackage{pifont}
\usepackage{xcolor}
\usepackage{listings}
\usepackage{amsmath}
\usepackage{tcolorbox}
\usepackage{hyperref}
\usepackage{tabularx}

\usepackage{amssymb}
\usepackage{pifont}
\usepackage[normalem]{ulem}

\newcommand{\cmark}{\checkmark}
\newcommand{\xmark}{\ding{55}}
\usepackage{xspace}
\newcommand{\mypara}[1]{\vspace{1.5pt}\noindent{\textbf{{#1}}\xspace}}
\newcommand{\myparait}[1]{\vspace{1.5pt}\noindent{\textit{{#1}}\xspace}}
\usepackage{enumitem}

\ifCLASSINFOpdf
\else
\fi
\begin{document}
%
\title{Technologies and Security Challenges in Metaverse}
%
%
%

\author{Krishno Dey,
        Diogo Barradas,
        and~Saqib Hakak
\thanks{K. Dey and S. Hakak are with the Canadian Institute for Cybersecurity, University of New Brunswick, NB, Canada, (e-mail: krishno.dey@unb.ca; saqib.hakak@unb.ca)}
\thanks{D. Barradas is with Cheriton School of Computer Science, University of Waterloo, ON, Canada, (e-mail: diogo.barradas@uwaterloo.ca)}
\thanks{Manuscript received April 19, 2005; revised August 26, 2015.}}

\markboth{IEEE Internet of Things Journal,~Vol.~14, No.~8, October~2025}%
{Shell \MakeLowercase{\textit{et al.}}: Bare Demo of IEEEtran.cls for IEEE Journals} 
%



\maketitle

\begin{abstract}
The Metaverse utilizes emerging technologies such as Extended Reality (XR), Artificial Intelligence (AI), blockchain, and digital twins to provide an immersive and interactive virtual experience. 
As the Metaverse continues to evolve, it brings a range of security and privacy threats, such as identity management, data governance, and user interactions. 
This survey aims to provide a comprehensive review of the enabling technologies for the Metaverse. It also aims to provide a thorough analysis of key vulnerabilities and threats that may compromise its sustainability and user safety.
We perform a systematic literature review (SLR) to identify key vulnerabilities and their countermeasures in Metaverse platforms. 
Metaverse offers a much larger attack surface compared to conventional digital platforms. Immersive, decentralized, and permanent characteristics of the Metaverse generate new vulnerabilities. Although there are many countermeasures to these vulnerabilities, most of them are theoretical or have not been tested in real-world environments.  
Our review highlights current advancements, identifies research gaps, and outlines future directions to ensure a secure, resilient, and ethically governed Metaverse.
\end{abstract}

\begin{IEEEkeywords}
metaverse, 
security threats, blockchain.
\end{IEEEkeywords}


%
\IEEEpeerreviewmaketitle

\section{Introduction}

The Metaverse is an immersive and persistent virtual platform that combines the physical world with the virtual world.  
It is still evolving and is considered the next-generation internet paradigm that will revolutionize the Internet. 
This virtual platform provides seamless social interaction, engagement, and entertainment within a self-sustaining ecosystem, while fostering rich user experiences \cite{meta-privacy-survey:wang2022survey}.
In contrast to conventional online platforms, the Metaverse seeks to go beyond simple two-dimensional interactions by offering highly realistic digital environments in which users can engage with a sense of near-physical presence and control. Technologies such as digital twin, extended reality (XR), artificial intelligence (AI), and blockchain pave the way for the development of such immersive and realistic environments \cite{ali2023metaverse}. The combination of these technologies empowers user interaction, content creation, and ownership in real time in the Metaverse platform. 
Future development of the Metaverse aims to provide an interoperable space for individuals and entities to engage in complex social and economic interactions, reflecting and enhancing real-world dynamics \cite{meta-privacy-survey:wang2022survey, he2022survey, wang2023survey}. Both industrial and scientific communities are converging to bring the vision of the future Metaverse to reality and transform daily digital interactions ranging from gaming, socializing, to collaborative work and e-commerce environments. 

The evolution of the Metaverse has been dependent on the development of various technologies that have progressively enabled and enriched the virtual interactive experiences. 
In early developments of the Metaverse, basic virtual reality (VR) systems, along with three-dimensional rendering-capable hardware, provided isolated immersive environments \cite{ali2023metaverse, ritterbusch2023defining}. These early VR systems were basic and had very limited paradigms for immersion and user interactions. However, they laid the foundation for the future development of the Metaverse.
Over time, the invention of augmented reality (AR) and mixed reality (MR) with VR developed into extended reality (XR), which expands the scope of sensory and interactive capabilities of the digital world. The emergence of digital twins helps produce a mirror image of the real world. 
Simultaneously, the emergence of blockchain technology provided decentralized digital asset verification, which improved the security in Metaverse platforms \cite{huynh2023blockchain}. The increased security offered by blockchain led to an exponential growth in economic transactions on Metaverse platforms and resulted in secure ownership of digital assets. 
The integration of artificial intelligence (AI) brought automation, adaptive user experiences, and advanced content generation. 
Moreover, advances in next-generation communication networks (notably 5G and the emergence of 6G) laid the groundwork for the low-latency and high-throughput data exchange necessary to sustain real-time, large-scale immersive environments \cite{chen2023metaverse}. 
Collectively, these expansions outline the development from isolated virtual environments to expansive, interconnected, and interoperable Metaverse ecosystems \cite{meta-privacy-survey:wang2022survey}.

The Metaverse has an expansive scope, covering a range of domains where its immersive and interactive capabilities are expected to have a significant impact. Prominent sectors include entertainment, healthcare, smart transportation, industrial manufacturing, education, and digital marketing \cite{meta-privacy-survey:wang2022survey, Raina2022MetaverseT}.
Virtual environments enable users to interact, play games, and engage socially with other users worldwide on online gaming platforms.
For example, healthcare applications use virtual environments for remote patient management, immersive rehabilitation therapies, and remote surgical supervision, leading to a significant transformation of traditional medical workflows \cite{Raina2022MetaverseT}.
The Metaverse facilitates safe virtual simulations that help with training, testing, and centralized real-time data analysis to improve operational efficiency in intelligent transportation systems.
Digital twin technologies in the Metaverse are used in industrial production to optimize product design, quality assurance, and testing, increasing flexibility and cutting expenses. The Metaverse is also utilized to offer a realistic representation of properties in digital real estate.
Furthermore, the Metaverse offers the potential to foster digital inclusion, bridging gaps for marginalized groups by creating innovative platforms for social and educational engagement, thus generating new socio-economic opportunities on a global scale.
Despite significant expansion in the Metaverse, security vulnerabilities remain the primary obstacles hindering its further development. The Metaverse presents a range of security challenges that hinder its safe and sustainable development \cite{meta-privacy-survey:wang2022survey, chen2023metaverse}. Issues such as identity theft, data breaches, avatar impersonation, and unauthorized access are becoming increasingly prevalent while users interact in immersive virtual environments. Malicious actors can take advantage of vulnerabilities in user authentication and data transmission due to the decentralized and persistent nature of the Metaverse. Furthermore, the integration of advanced technologies such as blockchain, artificial intelligence, and IoT devices poses further security challenges. Vulnerabilities of these technologies could easily be inherited by the Metaverse. Robust and adaptive security frameworks are required to address these security challenges and ensure confidentiality, integrity, and availability.


\begin{table*}[htbp]
\caption{A Comparison of Contribution Between Our Survey and Relevant Surveys}
\centering
\tiny
\begin{tabular}{@{}p{0.5cm}p{2.5cm}p{3cm}p{11cm}@{}}
\toprule
\textbf{Year} & \textbf{References} & \textbf{Focused Topic} & \textbf{Contribution} \\ \midrule

2008 & Leenes et al. \cite{leenes2007privacy} & Metaverse Security & Discusses privacy risks in the game Second Life from both social and legal perspectives. \\

2009 & Bourlakis et al. \cite{bourlakis2009retail} &  Metaverse Technology & Survey on Metaverse applications in terms of retailing. \\

2013 & Dionisio et al.\cite{dionisio20133d}  &  Metaverse Technology & Discusses key features of the Metaverse and ongoing improvements of the underlying virtual world technology. \\

2018 & Falchuk et al. \cite{falchuk2018social} & Metaverse Security & Survey on privacy issues and countermeasures related to digital footprints in social Metaverse games. \\

2020 & Diaz et al. \cite{diaz2020virtual} & Metaverse Application & Survey on applications of the Metaverse in education. \\

2021 & Duan et al. \cite{duan2021metaverse} & Metaverse Application & Survey on applications of the Metaverse in terms of social goods. \\

2021 & Lee et al.\cite{lee2024all} &  Metaverse Technology & Reviews fundamental technologies to develop the Metaverse. \\

2021 & Wang et al. \cite{wang2023survey} &  Metaverse Application  & Present the overview of Metaverse development in terms of national policies, industrial projects, infrastructures, supporting technologies, VR, and social metaverse. \\

2021 & Lee et al. \cite{lee2021creators} &  Metaverse Application & Survey on metaverse applications in terms of digital arts. \\

2022 & Yang et al. \cite{yang2022fusing} &  Metaverse Technology and Security & Discuss the potential of AI and blockchain technologies in future Metaverse construction. \\

2022 & Huynh-The et al. \cite{huynh2023artificial} & Metaverse Technology & Discuss the role of AI from six technical aspects in the development of the metaverse. \\

2022 & Park et al. \cite{park2022metaverse} &  Metaverse Technology & Discuss the hardware, software, and content components of the metaverse and review user interaction, implementation, and representative applications in the metaverse. \\

2022 & Xu et al. \cite{xu2022full} &  Metaverse Technology & An in-depth survey on the edge-enabled Metaverse in terms of communication, networking, and computation. \\

2022  & Wang et al. \cite{meta-privacy-survey:wang2022survey} &  Metaverse Technology and Security & Comprehensive survey of the fundamentals, security, and privacy of the metaverse. Discussions on security/privacy threats and state-of-the-art solutions, and future research directions in building a secure metaverse. \\

2023 & Ritterbusch et al. \cite{ritterbusch2023defining} & Metaverse Technology & Use a SLR to collect, analyze, and synthesize scientific definitions and the accompanying major characteristics of the Metaverse.\\

2023 & Huang et al. \cite{huang2023security} &  Metaverse Application and Security & Review key characteristics, applications of Metaverse, and highlight its critical security, privacy, and societal challenges.\\

2024 & Wang et al. \cite{wang2024integration} &  Metaverse Technology & Reviews the characteristics, architecture, and enabling technologies of Metaverse. Proposes Integrated Sensing, Communication, and Computing (SCC) to address resource challenges. \\ 

2025 & Huo et al. \cite{guo2025empirical} &  Metaverse Security &  Analyzes popular VR applications to identify widespread security threats and vulnerabilities.\\

2025 & Qin et al. \cite{qin2025identity} & Metaverse Security & Examine identity, crime, and law enforcement challenges in the Metaverse from a legal perspective and call for a unified international legal framework. \\

2025 & Zhang et al. \cite{zhang2025industrial} & Metaverse Technology & Explores enabling technologies, challenges, and future directions for the development of the industrial Metaverse.\\

2026  & Ours & Metaverse Technology and Security & \textit{(1) Conduct a systematic literature review on the fundamental technologies and security challenges in the Metaverse. (2) Present security threats and their countermeasures. }
\textit{(3) Proposed a structured taxonomy of threats and aligned them with the CIA triad. (4) Discuss future research directions in building a secure Metaverse.}\\ 
\bottomrule
\end{tabular}
\label{tab:survey-comparison}
\end{table*}

Table \ref{tab:survey-comparison} provides a comparison of the contribution of our work with other existing surveys. In this paper, we present a Systematic Literature Review (SLR) on the fundamental technologies and security challenges in Metaverse platforms. Our SLR provides readers with important insights and practical guidelines to help them better understand how these security/privacy issues could originate and be prevented in the Metaverse. The contributions of this survey are in the following:

\begin{itemize}
    \item \textit{We conduct a comprehensive Systematic Literature Review (SLR) on the fundamental technologies and security challenges in the Metaverse.} 
    \item \textit{We summarize the state-of-the-art (SOTA) technologies that are driving the development of Metaverse platforms.} 
    \item \textit{We identify and analyze common security threats to the Metaverse and their existing countermeasures in the literature.} 
    \item \textit{We propose a structured taxonomy of threats, aligning them with the CIA Triad}. 
    \item \textit{We highlight key research gaps and outline future directions for building more robust, secure, and trustworthy Metaverse platforms.}

\end{itemize}

The rest of the paper is organized as follows: Section \ref{sec:metaverse} provides a brief overview of the Metaverse.  Section \ref{sec:method} presents the research method used for this SLR. 
Section \ref{sec:rq1} highlights the Metaverse enabling technologies and their limitations. Then Section \ref{sec:taxonomy} categorizes security vulnerabilities of the Metaverse. Sections \ref{sec:rq3} and \ref{sec:rq4} present the security vulnerabilities and their countermeasures for Metaverse. Section \ref{sec:rq5} underscores the research challenges and future directions for the Metaverse. 
Following that, Section \ref{sec:limitations} identifies limitations of this study. Finally, Section \ref{sec:conclusion} concludes our findings. 



\section{Metaverse} \label{sec:metaverse}

\mypara{Definition and Evolution:} 
Metaverse is a three-dimensional online environment in which users represented by avatars interact with each other in virtual spaces decoupled from the real physical world \cite{ritterbusch2023defining}. The Metaverse is derived from the prefix ``meta'' (meaning transcendence) and the suffix ``verse'' (short for universe). The term was first used by Neal Stephenson in his science fiction novel Snow Crash in 1992 \cite{sfiction-book:stephenson1992snow}. Snow Crash introduced the term ``Metaverse'' to represent a parallel virtual world where humans can enter and live through digital avatars using virtual reality equipment. 
Early online games such as Habbo Hotel 
and Second Life 
offered foundational Metaverse experiences \cite{ritterbusch2023defining, chen2023metaverse}. They offered users interaction, creating virtual environments, and engaging in social activities within digital spaces. Later, online games such as Minecraft 
Roblox 
allowed users to generate content and offered more persistent and immersive digital world experiences \cite{chen2023metaverse}. The VR headset launched by Oculus\footnote{https://www.oculus.com/} marked a significant step toward advancing metaverse experiences, offering users immersive and interactive engagement within virtual environments. In recent times, Facebook (currently Meta) acquired Oculus and popularized the term Metaverse. 

\begin{figure}[!t]
\centering
\resizebox{\linewidth}{!}{
\includegraphics[scale=0.3]{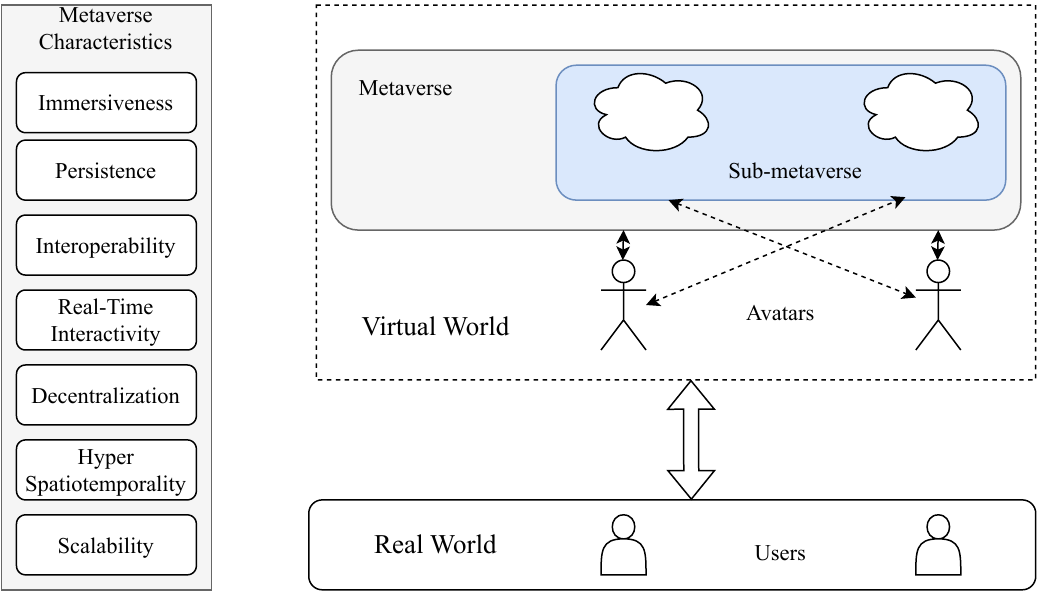}
}
\caption{Conceptual architecture of the Metaverse, highlighting key characteristics (e.g., immersion, interoperability, and decentralization).}
\label{fig:metaverse-characteristic}
\end{figure}

\mypara{Key Characteristics:}
Metaverse is an evolving paradigm of Web 3.0 technology, and it is essential to understand the fundamental characteristics that define its structure and functionality. In contrast to conventional digital platforms (in Web 1.0 and Web 2.0), the Metaverse offers a multidimensional, decentralized virtual environment. It utilizes advanced technologies to create immersive, persistent, and interactive experiences for the users. These unique features of the Metaverse are redefining the social, economic, and cultural interactions in digital worlds. The following are the core characteristics of the Metaverse. Figure \ref{fig:metaverse-characteristic} shows the basic architecture and highlights the characteristics of the Metaverse. 

\myparait{Immersiveness:} 
The Metaverse must be immersive to provide consumers with the best virtual experience \cite{meta-privacy-survey:wang2022survey, han2010user}. The term immersive means realistic, i.e., content in the Metaverse is sufficiently realistic for the users \cite{han2010user}. Immersive realism is another term that commonly refers to immersive Metaverse systems. Such realism in the virtual world allows users to sense their presence in virtual spaces. 
The Metaverse offers immersive environments through technologies such as virtual reality (VR) and augmented reality (AR), creating a sense of presence and realism.

\myparait{Persistence:}
The persistent nature of the metaverse means continuous existence, independent of the presence of individual users. In conventional digital platforms, the system is paused and restarted when no users are online \cite{han2010user}. In contrast, the metaverse's persistent nature means the virtual world continues to evolve and change even when users are offline, creating a living, breathing digital ecosystem \cite{han2010user, chen2023metaverse}. 

\myparait{Interoperability:}
Interoperability refers to the ability of different virtual environments to work together in a coordinated and seamless manner. Interoperability allows seamless interaction across different platforms, devices, and applications. It enables users to carry digital assets, identities, and data across virtual spaces and worlds \cite{yao2022metaverse}.

\myparait{Real-Time Interactivity:}
Real-time interactivity in the Metaverse refers to the ability of users to interact and experience a virtual world. Changes in the virtual world are reflected immediately in response to user actions and real-world events. Users in the Metaverse can also interact with each other in real-time. Such interaction allows dynamic social, economic, and collaborative experiences on virtual worlds \cite{yao2022metaverse, jaynes2003metaverse}.

\myparait{Decentralization:}
In the Metaverse, control and authority are distributed away from a central entity towards a network of users \cite{meta-privacy-survey:wang2022survey, chen2023metaverse}. Many Metaverse platforms utilize decentralized technologies such as blockchain to manage ownership, identity, and digital assets securely and transparently.

\myparait{Hyper Spatiotemporality:} The concept refers to a mode of existence or data processing that fundamentally breaks the traditional boundaries of physical time and space. 
 The irreversibility of time and the finiteness of space impose constraints on the real world. The Metaverse is a virtual world parallel to the real world and overcomes the constraints of time and space \cite{yao2022metaverse, huang2023security}. Hyper spatiotemporality allows users to move across various spatiotemporal dimensional worlds and provides a seamless alternate world experience. 

\myparait{Scalability:} 
The term "scalability" refers to the Metaverse's ability to maintain efficiency regardless of the number of users, data, and interactions \cite{chewslack2012scalable}. Scalability allows the system to expand to meet demand and contract when demand falls, ensuring profitability and a good user experience.


\mypara{Prototypes of Metaverse Applications}
Table \ref{tab:metaverse_prototypes} shows some of the popular existing prototypes of the Metaverse. We analyzed the documentation to identify different characteristics. 

\myparait{Games:}
Several existing game platforms serve as early prototypes of the Metaverse, showcasing its potential for immersive and interactive virtual experiences. Popular games such as Roblox, Fortnite\footnote{https://www.fortnite.com/}, and Decentraland\footnote{https://decentraland.org/} provide user-generated content, robust virtual economies, and rich social environments where users can create, explore, and interact. 
These gaming platforms demonstrate how digital worlds can function as more than just games, and could potentially evolve into social hubs and ecosystems \cite{han2021analysis}. 
Games such as Sandbox\footnote{https://www.sandbox.game/}, Axie Infinity\footnote{https://axieinfinity.com/}, and Minecraft incorporate features such as virtual land ownership, blockchain-based assets, and NFT integration, which further extend the concept of the Metaverse \cite{han2021analysis}. 
These platforms also allow users to monetize, trade digital goods, and engage in decentralized governance. They illustrate the fundamental components of the Metaverse (i.e., interoperability, user agency, and economic engagement) while highlighting the potential and difficulties associated with creating fully functional virtual environments.

\myparait{Socializing and Entertainment:}
Socializing and entertainment in the Metaverse are rapidly evolving through a range of prototypes that blend immersive technologies with interactive experiences. For example, platforms such as Second Life, VRChat\footnote{https://hello.vrchat.com/}, Roblox, Fortnite, Rec Room\footnote{https://www.therecroom.com/}, and Meta’s Horizon Worlds\footnote{https://horizon.meta.com/} demonstrate how users can engage in rich social interactions and virtual spaces \cite{paul2008socializing}. 
These platforms allow users to engage in activities such as 3D storytelling, avatar-based communication, and collaborative world-building. 
Metaverse is not limited to traditional gaming and transforms how events and entertainment are experienced. 
For example, virtual concerts featuring real-world artists have been hosted on platforms such as Fortnite, drawing millions of attendees \cite{chang2018mediation}. 
Similarly, institutions have started to utilize these virtual spaces to organize virtual graduation ceremonies and interactive art exhibitions, providing users with new ways to connect, learn, and celebrate in immersive environments \cite{chang2018mediation}. Moreover, digital museums are also growing in popularity, due to easier accessibility and cost-effectiveness.   
Aforementioned prototypes summarize the potential of the Metaverse to redefine social engagement and entertainment, while offering users globally accessible, customizable, and deeply interactive experiences.

\myparait{Workplace Collaboration:}
The Metaverse provides promising prototypes for immersive virtual collaboration, revolutionizing the way people work and communicate across distances. It enables real-time interactive engagement, making virtual workplaces, classrooms, and conference rooms increasingly viable.  
For example, Horizon Workrooms by Meta allows users from different physical locations to collaborate in a shared virtual office environment. Oculus Quest 2 headsets are used in Horizon Workrooms to have an immersive and interactive experience. Horizon Workrooms supports features such as spatial audio, gesture tracking, and virtual whiteboards, simulating in-person meetings. 
Similarly, Microsoft Mesh, an MR platform powered by Azure, uses holographic presence and shared virtual experiences to allow collaboration and enable teams to communicate in a digital replica of physical workspaces. 
These platforms demonstrate how the Metaverse may improve remote work, virtual meetings, panel discussions, and training sessions by providing users with a sense of presence, co-location, and involvement that goes beyond conventional video conferencing solutions \cite{darics2019talking}.
As these technologies continue to evolve, they are expected to play a central role in the future of hybrid and remote work.

\myparait{Simulation and Design:}
Simulation and design are significant areas of application of the Metaverse. It offers powerful tools for 3D modeling, visualization, and collaborative development. 
For instance, platforms such as NVIDIA Omniverse\footnote{https://www.nvidia.com/en-us/omniverse/} demonstrate the potential of the Metaverse by providing a real-time, multi-user virtual environment where users can simulate and design complex systems with high precision. 
Omniverse was originally developed for industrial use cases such as automotive design and architectural planning. It allows teams to work together in a shared virtual space and interact with representations of physical objects and environments. 
Omniverse is compatible with Universal Scene Description (USD), which is an open-source platform developed by Disney Pixar. It further enhances the utility of Omniverse by supporting interoperability across various design tools. 
These platforms of the Metaverse enable engineers, designers, and architects to collaborate remotely and test ideas in virtual prototypes,  while reducing the cost of physical resources and time of the design cycle.

\begin{table*}[ht]
\centering
\tiny
\caption{Summary of Existing Metaverse Prototypes in Different Applications}
\label{tab:metaverse_prototypes}
\begin{tabular}{llccccccc}
\toprule
\textbf{Prototype} & \textbf{Application} & \textbf{Immersive} & \textbf{Hyper Spatiotemporal} & \textbf{Open} & \textbf{Decentralized} & \textbf{Interoperable} & \textbf{Scalable} & \textbf{Heterogeneous} \\
\midrule
Second Life\footnote{https://secondlife.com/}           & MMO Game     & Partly & \cmark & Partly & \xmark & \xmark & \cmark & N/A \\
Roblox\footnote{https://www.roblox.com/}                & MMO Game     & \cmark & \cmark & \cmark & \xmark & Partly & \cmark & N/A \\
Fortnite\footnote{https://www.fortnite.com/}              & MMO Game     & \cmark & \cmark & Partly & \xmark & Partly & \cmark & N/A \\
Digital Palace Museum\footnote{https://intl.dpm.org.cn/index.html?l=en} & Travelling   & \cmark & \xmark & \xmark & \xmark & \xmark & Partly & N/A \\
Horizon Workroom\footnote{https://forwork.meta.com/ca/horizon-workrooms/}      & Working      & \cmark & \cmark & \cmark & \xmark & \xmark & Partly & N/A \\
Omniverse\footnote{https://www.nvidia.com/en-us/omniverse/}             & Simulation   & \cmark & \cmark & \cmark & \cmark & Partly & \cmark & \cmark \\
Decentraland\footnote{https://decentraland.org/}          & Game         & \cmark & \cmark & \cmark & \cmark & \xmark & \cmark & Partly \\
Cryptovoxels\footnote{https://www.cryptovoxels.com/}          & Game         & \cmark & \cmark & \cmark & \cmark & \xmark & \cmark & Partly \\
Minecraft\footnote{https://www.minecraft.net/en-us}             & Sandbox Game & Partly & \cmark & \cmark & \xmark & Partly & \cmark & N/A \\
VRChat\footnote{https://hello.vrchat.com/}                & Social VR    & \cmark & Partly & \cmark & \xmark & \xmark & Partly & Partly \\
Rec Room\footnote{https://recroom.com/}              & Social VR    & \cmark & Partly & \xmark & \xmark & \xmark & Partly & N/A \\
The Sandbox\footnote{https://www.sandbox.game/en/}           & Game         & \cmark & \cmark & \cmark & \cmark & \xmark & \cmark & Partly \\
Spatial\footnote{https://www.spatial.io/}               & Collaboration & \cmark & Partly & \cmark & \xmark & Partly & Partly & Partly \\
NVIDIA RTX World\footnote{https://www.nvidia.com/en-us/geforce/rtx/} & Simulation   & \cmark & \cmark & \xmark & \xmark & \xmark & \cmark & \cmark \\
Blue Mars\footnote{https://www.mmogames.com/game/blue-mars}             & MMO Game     & \cmark & Partly & \xmark & \xmark & \xmark & Partly & N/A \\
Somnium Space\footnote{https://www.somniumspace.com/}         & Game/VR      & \cmark & \cmark & \cmark & \cmark & \xmark & \cmark & Partly \\
Meta Quest Home\footnote{https://www.meta.com/help/quest/811926333306947/?srsltid=AfmBOopJ3JqdDxotGrqz4fyjSVBeephrxLPQ-izLD8S53HwGU0OxFa8V}       & Social VR    & \cmark & \xmark & \xmark & \xmark & \xmark & \xmark & N/A \\
Neos VR\footnote{https://neos.com/}               & Collaborative VR & \cmark & \cmark & \cmark & \cmark & Partly & \cmark & Partly \\
Mozilla Hubs\footnote{https://support.mozilla.org/en-US/kb/end-support-mozilla-hubs}          & Web-based VR & \cmark & \xmark & \cmark & \xmark & \cmark & Partly & Partly \\
High Fidelity\footnote{https://store.steampowered.com/app/390540/agecheck}         & Social Audio & Partly & \xmark & \xmark & \xmark & Partly & Partly & N/A \\

\bottomrule
\end{tabular}
\end{table*}

\section{Methodology} \label{sec:method}
In this section, we present the research methodology applied to this study. First, we outline the selected research questions. Next, we provide a detailed explanation of our research strategy, including the search string, literature resources, and the article search process. Following that, we describe how we selected articles for the study. Finally, we demonstrate the data synthesis from those articles.

\mypara{Research Questions:}
\begin{itemize}
        \item \textit{RQ1: What are the fundamental technologies that are driving the development of current Metaverse platforms?}
        \item \textit{RQ2: What are the major security threats and vulnerabilities in Metaverse platforms?}
        \item \textit{RQ3: What defense mechanisms or technologies are used to address security threats in the Metaverse?}
        \item \textit{RQ4: What are the open research challenges in ensuring secure and trustworthy Metaverse environments?}
\end{itemize}

\mypara{Research Strategy:}
The steps involved in creating a search term, creating a search string, and identifying the literature sources used for this research are detailed below.

\myparait{Search Stings:}
To construct the search string, we adhered to the universally accepted systematic literature guidelines developed by Kitchenham \cite{kitchenham2007guidelines}. The following steps were undertaken to create an appropriate search string.

\begin{enumerate}
    \item Derivation of major terms: Commence by identifying the central terms associated with Metaverse.
    \item Identification of alternative spellings and synonyms: Locate synonyms or related terms for "Metaverse."
    \item Identification of keywords in relevant papers or books: Explore existing literature, papers, or books related to Metaverse to identify keywords utilized in those publications.
    \item Usage of Boolean operators (OR and AND): Integrate these terms using Boolean operators to craft a comprehensive search string.
\end{enumerate}

After completing all the aforementioned steps, we have formulated the following search string.

("Metaverse" OR "Virtual Reality" OR "Augmented Reality" OR "Extended Reality" OR "XR") 
AND ("security" OR "cybersecurity" OR "privacy" OR "data protection" OR "threats" OR "attacks" OR "vulnerabilities") 
AND ("technology" OR "architecture" OR "framework" OR "blockchain" OR "AI" OR "machine learning" OR "IoT" OR "digital twin" OR "AIGC" OR "6G" OR "edge computing")

\myparait{Literature Resources:}
We employed Scopus, IEEE Xplore, ACM, USENIX, ScienceDirect, Web of Science, and Google Scholar to discover relevant papers for our review study. To search published journal papers, conference proceedings, and IEEE bulletins, we utilized titles, abstracts, and keywords.

\mypara{Selected Studies:}
In this section, a comprehensive set of rigorous scrutiny steps was undertaken to further synthesize the initially selected articles. The detailed scrutiny process is outlined below.

\myparait{Scrutiny:}
The initial search process resulted in the identification of a large number of potential studies related to Metaverse enabling technologies and security challenges. Consequently, comprehensive scrutiny was essential to narrow down the studies. During this scrutiny, we initially considered the titles and brief summaries of the articles. We began by filtering articles based on whether they were written and published in English, considering sources such as journals, conference proceedings, and IEEE bulletins. Papers that did not address any of the research questions or did not align with the discussion topic were excluded. In instances where multiple copies of the same article were found on the internet, we selected the most comprehensive version. Additionally, our SLR focused on articles published between January 1, 2008, and July  30, 2025. The table \ref{table:inclusion_exclusion_criteria} illustrates the inclusion and exclusion criteria employed during the scrutiny process. We narrowed down our selection to 85 papers overall after carefully following the instructions provided.

\begin{table}[!ht]
\centering
\tiny
\caption{Inclusion and Exclusion Criteria}
\label{table:inclusion_exclusion_criteria}
    \begin{tabular}{p{0.45\linewidth}p{0.45\linewidth}}
    \toprule
        \textbf{Inclusion criteria} & \textbf{Exclusion criteria} \\
        \midrule
        a. All English-language papers & a. Non-English published papers \\
        b. Papers focusing on security in Metaverse & b. Papers unrelated to the research questions \\
        c. Relevant papers published from 2008 to 2025 & c. Papers lacking essential bibliographic information like publication date, type, volume, or issue numbers \\
        d. All published papers capable of addressing at least one research question & d. Only the most comprehensive, recent, and improved version of duplicate papers is included, excluding the others. \\
        \bottomrule
    \end{tabular}
\end{table}

\mypara{Data Synthesis:}
To adequately address the research questions, a thorough data synthesis process was used in this section to extract insights from the chosen studies. The selected studies were thoroughly examined, and their contents were evaluated.
The main goal of this effort was to bring the chosen studies into harmony, which would enhance comprehension and make it easier to pinpoint exact responses to the revised research questions. The data that was extracted included both quantitative components and qualitative components. The following describes the methodology used to carry out data synthesis, which is customized for every research question: 

\begin{itemize}
    \item To answer RQ1, we use architectural figures to present the enabling Metaverse technologies. We use descriptive analysis to highlight the limitations of current Metaverse technologies. 


    \item To answer RQ2 and RQ3, we analyze the existing literature to identify security threats and their corresponding countermeasures in the Metaverse.

    \item Finally, for RQ4, we identify the gaps from current research and propose future research directions.
\end{itemize}

\section{Fundamental Technologies for Metaverse and their limitations (RQ1)} \label{sec:rq1}

\begin{figure}[!t]
\centering
\resizebox{0.5\linewidth}{!}{
\includegraphics[scale=0.35]{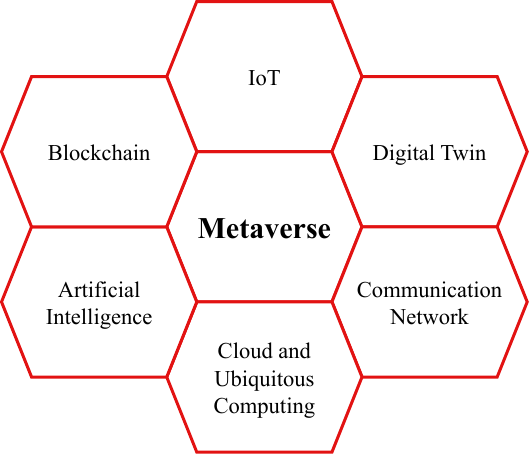}
}
\caption{Enabling technologies of Metaverse. }
\label{fig:metaverse-technologies}
\end{figure}

The Metaverse is a persistent, immersive, and interactive 3D digital environment. 
It is built upon the convergence of several foundational technologies and allows users to engage socially, economically, and creatively. 
These core technologies not only help in the development of the Metaverse but also help make the Metaverse scalable, secure, and usable. Figure \ref{fig:metaverse-technologies} shows all the major enabling technologies of the Meataverse.
Extended Reality (XR), Artificial Intelligence (AI), Blockchain, the Internet of Things (IoT), Digital Twins, Cloud and Edge Computing, 5G/6G communication networks, and Software-Defined Networking (SDN) are key technologies for enabling the Metaverse \cite{meta-privacy-survey:wang2022survey, chen2023metaverse}.

Despite the rapid development and innovation of the Metaverse ecosystem, existing technologies contain significant drawbacks and challenges. These limitations could potentially hinder broad adoption, seamless user experiences, and system-level efficiency of the Metaverse. These challenges span across the technical, infrastructural, social, and economic domains. Despite the enormous potential of foundational technologies such as XR, AI, blockchain, IoT, and 5G/6G, problems with scalability, latency, security, standardization, user experience, cost, and energy consumption plague their existing implementations.

\mypara{Extended Reality (XR):} comprises of Virtual Reality (VR), Augmented Reality (AR), and Mixed Reality (MR), and is responsible for the immersive nature of the Metaverse \cite{meta-privacy-survey:wang2022survey}. 
AR overlays digital content onto the physical world, VR creates computer-generated 3D synthetic environments, and MR combines the two by providing real-time interaction between digital and physical worlds. These technologies enable users to experience an immersive sense of “presence” within the Metaverse, simulating real-world environments.
To enhance this immersive experience, wearable devices such as Head-mounted displays (HMDs), haptic gloves, motion trackers, and spatial audio systems are commonly used. 
Wearable XR devices continue to evolve and support personalized and dynamic virtual interactions. These devices also incorporate ubiquitous sensing capabilities to capture user behavioral data. 

XR has several limitations, including device standardization, data quality, and privacy controls. Many sensitive pieces of information, including eye movements, gestures, and spatial orientation, are gathered by headsets and sensors and may be stored or transferred insecurely.  
These systems often lack robust authentication mechanisms \cite{zhang2024anti}, leading to threats such as identity firmware exploits \cite{zhang2025industrial}, biometric data leaks \cite{huang2023security}, impersonation \cite{brown2018security}, session hijacking \cite{ryu2022design}, and spoofing attacks \cite{mitrushchenkova2023personal}. In addition, malicious virtual content can be used for social engineering, where fake avatars or manipulated environments deceive users into revealing private information or taking unsafe actions.

\mypara{Communication Networks:}
The Metaverse demands ultra-fast, low-latency, high-bandwidth connectivity, which is supported by 5G networks and will be significantly enhanced by 6G in the near future \cite{du2022optimal, cheng2024user}. 5G enables real-time multiplayer interactions, seamless streaming of HD virtual environments, and efficient device-to-device communication. With even higher data rates and lower latency, 6G is expected to support AI-native networking, context-aware services, and integrated communication between XR devices, IoT sensors, and cloud platforms. These networks are foundational for maintaining synchronous and immersive experiences across geographically dispersed users and devices.
SDN (Software-Defined Networking) is a network architectural technique that makes it possible to govern network traffic in a centralized, programmable manner \cite{meta-privacy-survey:wang2022survey}.  SDN aids in the dynamic allocation of network resources within the Metaverse to preserve the quality of service (QoS) for latency-sensitive applications like real-time simulations, VR gaming, and remote collaboration.  Through the abstraction of the control plane from the hardware architecture, SDN offers effective traffic management, scalability, and flexibility.  Supporting millions of concurrent users in virtual environments requires the ability to prioritize Metaverse services and quickly adjust to changing network needs.

Due to the increased granularity of 5G and 6G, they are vulnerable to security threats such as location monitoring \cite{yang2023secure}, rogue base stations \cite{de2023survey}, and network slicing vulnerabilities \cite{de2023survey}. They are also prone to configuration errors and interception attacks \cite{yang2023secure}, such as man-in-the-middle (MitM) attacks \cite{ryu2022design}, due to their sophisticated infrastructure. Furthermore, in the absence of strong access control, these networks may enable cross-layer attacks and compromise the confidentiality and integrity of data. 
The centralized network control in SDN brings in the risk of a single point of failure \cite{wang2021blockchain}. Attackers can target the SDN controller to perform flow rule manipulation, network reconnaissance, or even complete network takeover \cite{du2022optimal}. Further limitations include vulnerabilities in east-west communication that might cause lateral movement across virtualized systems, controller overload, and weak authentication \cite{yang2023secure}. 

\mypara{Digital Twin (DT):} are digital copies of real-world objects, such as people, systems, or processes. It helps generate a realistic replica of the physical world in the Metaverse \cite{wu2021digital}. Digital twins are widely used in domains such as healthcare, manufacturing, urban planning, and automotive design. Digital twin enables collaborative real-time 3D simulation on platforms such as NVIDIA's Omniverse, allowing professionals to interact and refine physical designs in a virtual space \cite{liao2021digital}. Digital twins facilitate predictive analysis, optimization, and visualization by incorporating real-time data from Internet of Things (IoT) devices. This enhances operational efficiency and decision-making in virtual settings that replicate real-world conditions.

DT heavily relies on real-time and accurate data for replication. If compromised data is used for replication, it can lead to incorrect decision-making or false representation.
Attackers can manipulate the digital twin to impersonate users \cite{brown2018security}, conduct data injection attacks \cite{liang20162015}, exploit unauthorized access \cite{siddiqi2024multichain} to sensitive mirrored environments, and manipulate DT simulation \cite{kuru2023metaomnicity}. A corrupted DT can also have serious physical-world consequences if it is connected to control systems (e.g., in industrial settings). 

\mypara{Artificial Intelligence (AI):} enables smart interactivity and automation within Metaverse environments.  It generates realistic avatars, non-player characters (NPCs), adaptive environments, voice recognition systems, and real-time language translation \cite{chen2023metaverse}. AI is also used to predict user behavior and recommend appropriate content to users. It also helps in digital asset creation and management in the digital world. Generative AI is used to generate avatars, graphics, music, and narrative in virtual spaces. In addition, AI is also utilized to identify anomalous behaviors, detect security threats, and enable personalized access control mechanisms.  While machine learning (ML) and deep learning (DL) models improve simulation, interactivity, and personalization, reinforcement learning facilitates autonomous decision-making in the Metaverse.

Explainability problems, bias in training data, and the possibility of generative model abuse are major concerns and limit the use of AI. 
Attackers may exploit these weaknesses for adversarial attacks \cite{adversarial}, where small perturbations cause misclassifications in AI decisions. Additionally, identity theft \cite{mitrushchenkova2023personal} can be committed with the use of AI-generated deepfakes \cite{tariq2023deepfake}, and model corruption during training can be caused by data poisoning \cite{adversarial} assaults. 
AI systems may also unintentionally compromise privacy by profiling user behavior beyond ethical or legal boundaries.

\mypara{Blockchain:} provides decentralized control, transparency, and security within the Metaverse. It allows us to create and exchange digital assets such as Non-Fungible Tokens (NFTs), which provide verifiable ownership of virtual commodities such as avatars, apparel, land, and collectibles \cite{gai2022blockchain}. Additionally, blockchain provides decentralized identity systems (DID) and enables users to safely manage their identities across several platforms. Smart contracts automate transactions and provide secure governance, virtual economies, and decentralized application (dApp) operations \cite{sayeed2020smart}. Blockchain enhances user trust and interoperability in multi-platform Metaverse ecosystems, as it uses both public and consortium networks to manage authentication, traceability, and secure data sharing.

Blockchain utilizes NFTs (Non-Fungible Tokens) and virtual currency to provide a secure virtual economy. 
However, the immutability property of blockchain can have several drawbacks, making it challenging to reverse fraudulent transactions. 
Unsecurely coded smart contracts can be exploited through reentrancy attacks \cite{alkhalifah2021mechanism} and logic bugs \cite{sun2024gptscan, wang2020oracle}. Furthermore, blockchain pseudonymity can result in fraudulent identity masking or money laundering \cite{nawari2019blockchain}. 
Blockchain systems also face latency, throughput, scalability, and energy consumption challenges \cite{nawari2019blockchain}.

\mypara{Internet of Things (IoT):} makes up the Metaverse's sensory and data collection layer. It allows for real-time data streaming and contextual responsiveness by connecting physical devices, sensors, and actuators to virtual platforms. For instance, wearable devices, biometric sensors, smart cameras, and environmental monitors all continuously send data into the Metaverse, enabling digital avatars and systems to respond to changes in the real world \cite{jaynes2003metaverse}. IoT and digital twins work together to assist industrial applications in replicating real-time urban or manufacturing environments. This integration further improves the Metaverse's realism and interactivity by enabling ubiquitous computing, in which digital services adapt fluidly to user context.

IoT devices are resource-constrained, which makes it difficult to implement strong encryption or regular updates. Absence of strong encryption on these devices could lead to firmware vulnerabilities and insecure communication channels. Common threats include device hijacking \cite{ometov2016facilitating, liu2009botnet}, sensor spoofing \cite{ali2023metaverse}, and insecure IoT APIs \cite{huang2023security}. Furthermore, Attack surfaces expand with the size of IoT networks, making it feasible to launch lateral movement attacks from a single compromised device.

\mypara{Ubiquitous Computing:} The term "ubiquitous computing," sometimes referred to as "pervasive computing," refers to the seamless integration of computing capabilities into everyday environments, allowing systems and devices to run continually and covertly in the background \cite{vural2012survey}. Ubiquitous computing is essential to the Metaverse's development of intelligent, contextually aware settings that facilitate real-time interaction between users, digital systems, and physical objects. 
The widespread use of wearables, sensors, smart devices, and embedded systems makes it possible to continuously collect data and process the data, resulting in personalized and adaptive experiences. For instance, smart glasses can provide real-time information overlays, and connected home devices can respond to user preferences without requiring direct input. This invisible network of computing resources ensures that users remain continuously connected to the Metaverse, enhancing immersion and enabling seamless transitions between physical and virtual worlds. As ubiquitous computing evolves, it will be fundamental to achieving the vision of a persistent, intelligent, and user-centric Metaverse.

Privacy and surveillance problems are introduced by the ubiquitous computing, ambient intelligence, and always-on nature.  These systems often operate with minimal user awareness, and attackers can launch passive surveillance \cite{falchuk2018social}, sensor spoofing \cite{mitrushchenkova2023personal}, or unauthorized data collection \cite{ometov2016facilitating}. Other limitations include a lack of user control, inconsistent standards, and difficulty in auditing distributed systems, leading to data leakage and context manipulation.

\mypara{Cloud and Edge Computing:} infrastructures provide the significant processing power needed to render realistic 3D environments, manage enormous datasets, and ensure real-time responsiveness \cite{weimann2023metaverse}.  Cloud computing supports scalability and global accessibility by offering centralized resources for massive data processing, simulation, and storage.  In contrast, edge computing reduces latency and allows for real-time interaction, bringing computation closer to the user. This is crucial for responsive collaboration in the Metaverse. 
By dynamically allocating rendering tasks across client and cloud devices, platforms such as CloudRend balance resource consumption and enhance visual performance.

Despite having the scalability of cloud computing, they suffer from multi-tenancy risks, latency issues, and a lack of transparency in data management. These systems are susceptible to insider threats, data breaches \cite{wei2020ldp}, and Denial-of-Service (DoS) \cite{bertino2017botnets} attacks. In the Metaverse, edge nodes can be targeted for physical tampering, data manipulation, or unauthorized access.


 \begin{tcolorbox}[boxrule=0.5pt,boxsep=2pt,left=1pt,right=1pt,top=1pt,bottom=1pt]
\noindent\textbf{RQ1 Summary:} The Metaverse is the result of several cutting-edge technologies coming together and cooperating, rather than a single technological innovation.  
These technologies work together as the foundation of modern Metaverse platforms, enabling a wide range of applications in fields such as business, education, healthcare, gaming, and entertainment. 
Table \ref{tab:rq2-tech-threats} provides a summary of the security threats associated with the use of metaverse technologies. The complexity of the metaverse is further increased by the high expenses and moral dilemmas associated with governance and content moderation.  It will take interdisciplinary cooperation and the development of unified standards and protocols to close these gaps.  

\end{tcolorbox}

\begin{table*}[]
\centering
\tiny
\renewcommand{\arraystretch}{1.2} 
\caption{Threat Categories and Their Aligned Technologies}
\label{tab:rq2-tech-threats}
\begin{tabular}{p{5cm} p{9cm}}
\toprule
\textbf{Technologies} & \textbf{Aligned Threats} \\ \hline

XR & Firmware Exploits and  Biometric Data Leaks \\

Networks & SPoF, DDoS Attack, Sybil Attack, Man-in-the-middle (MitM) Attacks, Jamming of 5G/6G Signals, SDN Controller Attacks, Network Slicing Vulnerabilities\\

DT & Unauthorized Access to DT Models,  Manipulation of DT Simulations\\

AI & False Data Injection, Malicious Bots and NPCs \\ 

Blockchain & Smart Contract Bugs, Private Key Theft, Fraudulent NFTs\\

IoT & Compromised Devices, Sensor Spoofing, Insecure Device APIs\\

Cloud \& Ubiquitous Computing & Data Leakage, Insider Threats \\ 

\bottomrule
\end{tabular}
\end{table*}
\section{Taxonomy of Threats on Metaverse}  \label{sec:taxonomy}
\begin{figure*}[!t]
\centering
\resizebox{\linewidth}{!}{
\includegraphics[scale=0.5]{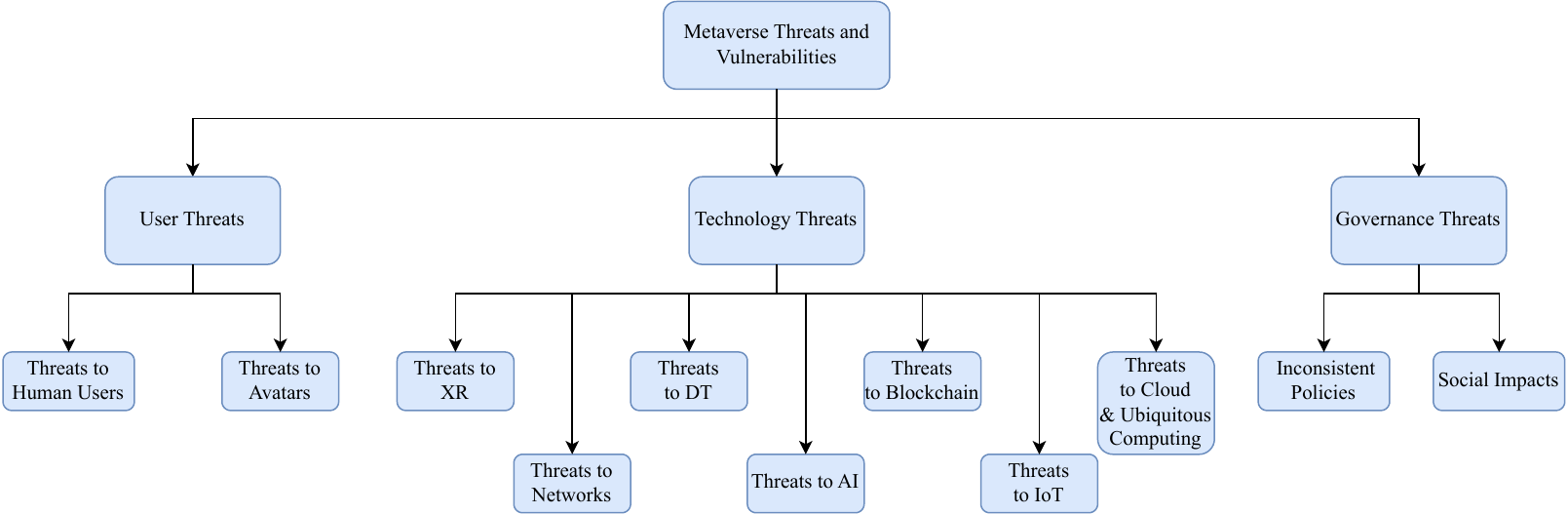}
}
\caption{Taxonomy of Threats in the Metaverse.}
\label{fig:metaverse-taxonomy}
\end{figure*}

We propose a thorough taxonomy of threats that aligns each identified risk with the well-known CIA triad (Confidentiality, Integrity, and Availability). 
Figure \ref{fig:metaverse-taxonomy} classifies threats according to several categories and sub-categories. Table \ref{tab:metaverse-cia-threats-user}, \ref{tab:metaverse-cia-threats-technology}, and \ref{tab:metaverse-cia-threats-governance} aligns the classified threats with the CIA triad. Such an organized approach makes it easier to comprehend how distinct threat vectors compromise particular security principles and the consequences these have on the reliability and functionality of Metaverse systems.
We divide threats related to the Metaverse users into human users and avatars. Human users are prone to identity theft, impersonation, inadequate multi-factor authentication, and pervasive data collection. 
In contrast, threats related to avatars comprise authentication, unauthorized access to virtual spaces, and malicious virtual assets. Furthermore, threats to virtual trading, asset ownership, and the economy are also alarming problems. These undermine both confidentiality and integrity, often leading to unauthorized access or the misuse of user identities. 

The category of Technological threats comprises threats such as threats related to Extended Reality (XR), Networks, Digital Twin (DT), Artificial Intelligence (AI), Blockchain, Internet of Things (IoT), and Cloud and Ubiquitous Computing. 
Finally, governance threats and challenges highlight the legal and regulatory challenges in upholding law and order on Metaverse platforms. There are significant gaps in availability and integrity as a result of unclear virtual crime legislation, regulatory power abuse, and a lack of cooperative governance structures. 

Overall, this structured taxonomy provides a fundamental framework for assessing and resolving the complex security issues that exist in the Metaverse. By mapping threats to core security principles such as CIA, we not only highlight the technical dimensions of each risk but also emphasize their broader implications for user trust, operational reliability, and long-term sustainability. Additionally, it gives researchers, developers, and politicians a strategic prism through which to prioritize defense mechanisms and create Metaverse systems that are secure by default. In Section \ref{sec:rq3} we discuss all the threats in more detail.

\begin{table*}[t]
\centering
\tiny
\caption{Metaverse Threat Taxonomy based on User Threats}
\label{tab:metaverse-cia-threats-user}

\begin{tabular}{@{}p{2.5cm}p{4.5cm}p{5.5cm}p{2.5cm}@{}}
\toprule
\textbf{Users} & \textbf{Category} & \textbf{Specific Threat} & \textbf{CIA Component} \\
\midrule

\multirow{4}{*}{Human} 
& Identity Abuse & Identity theft \cite{mitrushchenkova2023personal}, impersonation attacks \cite{brown2018security}, session hijacking \cite{mitrushchenkova2023personal}, deepfake-based impersonation \cite{tariq2023deepfake}  & Confidentiality, Integrity \\
& Authentication Weakness & Inadequate multi-factor authentication \cite{yang2023secure} & Confidentiality \\
& Privacy Risks & Extensive data collection \cite{shang2020arspy},  emerging data modalities \cite{cheng2024user}, risks to user safety \cite{casey2019immersive} & Confidentiality, Integrity \\
& Social Engineering & Phishing \cite{manoharan2025metahuman},  data manipulation attacks \cite{su2020lvbs}, and behavioral manipulation \cite{ali2023metaverse} & Confidentiality, Integrity \\
\midrule

\multirow{6}{*}{Avatar} 
& Avatar Authentication & Weak avatar authentication mechanisms  \cite{yang2023secure, tariq2023deepfake} & Confidentiality \\
& Access Control & Unauthorized access to virtual spaces \cite{xu2022wireless, yu2018leveraging} & Confidentiality \\
& Asset Integrity   & Malicious virtual assets \cite{hussain2023impact, oh2024enhances} & Integrity \\
& Virtual Economy & Threats to virtual trading systems \cite{liao2021digital, wanick2023brand} & Integrity, Availability \\
& Ownership Security & Threats to digital asset ownership \cite{wang2021non, popescu2021non}  & Confidentiality, Integrity \\
& Economic Manipulation & Threats to digital economy \cite{zhang2021privacy, xu2014collusion} & Integrity \\
\bottomrule

\end{tabular}
\end{table*}

\begin{table*}[t]
\centering
\tiny
\caption{Metaverse Threat Taxonomy Mapped Based on Technology}
\label{tab:metaverse-cia-threats-technology}

\begin{tabular}{@{}p{3cm}p{6cm}p{3cm}@{}}
\toprule
\textbf{Category} & \textbf{Specific Threat} & \textbf{CIA Component} \\
\midrule

\multirow{2}{*}{XR} 
& XR firmware exploits \cite{zhang2025industrial, jaber2022security} & Confidentiality, Integrity \\
& Biometric data leakage \cite{huang2023security} & Confidentiality \\
\cmidrule{1-3}

\multirow{7}{*}{Network} 
& Single Point of Failure attacks \cite{wang2021blockchain, noveck2011single} & Availability \\
& Distributed Denial-of-Service attacks \cite{bertino2017botnets, kuo2023metaverse} & Availability \\
& Sybil attacks \cite{cao2024eden, kim2022novel} & Confidentiality, Integrity \\
& Man-in-the-Middle attacks \cite{su2021secure, vondravcek2023rise} & Confidentiality, Integrity \\
& Jamming attacks \cite{yang2023secure, ali2023metaverse} & Availability \\
& SDN controller attacks \cite{wang2021blockchain} & Integrity, Availability \\
& Network slice isolation failures \cite{de2023survey} & Confidentiality, Integrity \\
\cmidrule{1-3}

\multirow{2}{*}{Digital Twin} 
& Unauthorized access to digital twin models \cite{yao2022metaverse} & Confidentiality, Integrity \\
& Manipulation of digital twin simulations \cite{kuru2023metaomnicity,  liao2021digital} & Integrity \\
\cmidrule{1-3}

\multirow{2}{*}{AI} 
& Data injection attack \cite{liang20162015, siddiqi2024multichain} & Integrity \\
& Malicious autonomous agents \cite{han2022cheating} & Integrity \\
\cmidrule{1-3}

\multirow{3}{*}{Blockchain} 
& Smart contract exploitation \cite{brown2018security, alkhalifah2021mechanism} & Integrity, Availability \\
& Private key theft \cite{ryu2022design} & Confidentiality \\
& Fraudulent digital assets \cite{popescu2021non, wang2021non} & Integrity \\
\cmidrule{1-3}

\multirow{3}{*}{IoT} 
& Compromised devices \cite{blow2020study, shang2020arspy} & Confidentiality, Integrity \\
& Sensor spoofing attacks  \cite{ali2023metaverse, chen2023metaverse} & Integrity, Availability \\
& Insecure IoT interfaces \cite{ali2023metaverse, huang2023security} & Confidentiality, Integrity \\
\cmidrule{1-3}

\multirow{2}{*}{Cloud \& Ubiquitous Computing} 
& Data leakage \cite{li2021verifiable, gdprinfoGeneralData} & Confidentiality \\
& Insider threats  \cite{wei2020ldp, bertino2017botnets} & Confidentiality, Integrity \\

\bottomrule
\end{tabular}
\end{table*}

\begin{table*}[htbp]
\centering
\tiny
\caption{Metaverse Threat Taxonomy based on Governance}
\label{tab:metaverse-cia-threats-governance}

\begin{tabular}{@{}p{3cm}p{6cm}p{3cm}@{}}
\toprule
\textbf{Category} & \textbf{Specific Threat/Challenges} & \textbf{CIA Component} \\
\midrule

\multirow{3}{*}{Governance and Policy}
& Lack of legal frameworks for virtual crimes \cite{hendaoui20083d, qin2025identity} & Integrity, Availability \\
& Regulatory misconduct \cite{bai2021public, sayeed2020smart} & - \\
& Policy inconsistency \cite{huang2019software} & - \\
\cmidrule{1-3}

\multirow{4}{*}{Social and Systemic Impacts}
& Digital forensics limitations \cite{li2021toward, tariq2023deepfake} & Integrity, Confidentiality \\
& Infrastructure safety risks \cite{vellaithurai2014cpindex} & Availability \\
& Intellectual property violations \cite{hendaoui20083d, btljRightPublicity} & Confidentiality, Integrity \\
& Adverse social impacts \cite{valluripally2021modeling, leenes2007privacy, zhu2020activity} & - \\

\bottomrule
\end{tabular}
\end{table*}

\section{Security Threats and Vulnerabilities in Metaverse (RQ2)}  \label{sec:rq3}
The Metaverse offers an immersive and persistent digital ecosystem. However, the Metaverse is subject to a wide array of novel and intensified security threats. In contrast to conventional web-based systems, the Metaverse includes real-time interactions in three-dimensional environments, complex identity representations through avatars, and the incorporation of cutting-edge technologies such as XR, blockchain, AI, and IoT. Although these features provide for immersive experiences, they can amplify preexisting weaknesses and open new attack surfaces. The more the Metaverse develops, the more important it is to comprehend the range of security risks it faces.

\subsection{Metaverse Users}
The Metaverse's real-time and multi-platform nature makes secure and seamless authentication essential. However, current authentication methods are often inadequate \cite{yang2023secure}. Impersonation, replay, and man-in-the-middle (MITM) attacks can result from weak credential systems, a lack of multi-factor authentication, and unsafe session handling \cite{ryu2022design}. Platforms such as VRChat and Horizon Worlds do not always include strong identification verifications, making them vulnerable to these kinds of attacks. 

\subsubsection{Human Users}
In the Metaverse, human users refer to the core participants actively involved in the environment. Human users are susceptible to identity theft, impersonation, social engineering, and privacy violations attacks.

\paragraph{Identity Abuse} 
Exposure of a user's identity in the Metaverse can have serious consequences. The consequences of such exposure can be far more severe than traditional systems, resulting in the loss of their avatars, digital assets, social connections, and even their entire virtual presence \cite{mitrushchenkova2023personal}. For instance, in virtual platforms, attackers can exploit vulnerabilities such as compromised VR headsets, phishing scams, or weaknesses in authentication mechanisms. 

An impersonation attack involves cybercriminals posing as trusted individuals or organizations \cite{brown2018security}. Such a disguise helps attackers to gain confidential information, transfer funds, or grant access to secure systems \cite{barbeau2005detecting}. To perform impersonation attacks, cybercriminals employ social engineering techniques \cite{hu2015dynamic}. 

\begin{enumerate}
    \renewcommand{\labelenumi}{\roman{enumi}.}
    \item Session Hijacking: 
    Session hijacking in the metaverse refers to an attacker taking control of an active session between a user and a service to obtain unauthorized access.
    Attackers can bypass authentication and obtain control of the avatar, access personal data, or manipulate transactions if session tokens or cookies are stolen. The integration of XR devices and IoT sensors further amplifies the session hijacking threat. Weak session management or insecure transmission channels could expose tokens to interception through man-in-the-middle (MitM) attacks, replay attacks, or malware \cite{mitrushchenkova2023personal}. Ensuring secure session handling, encryption, and continuous re-authentication is essential for the security of the Metaverse platforms.

    \item Deepfake-Based Impersonation: 
    Deepfake technologies leverage generative adversarial networks (GANs) to fabricate hyper-realistic digital representations of legitimate users, undermining traditional authentication protocols. These AI-driven fabrications can be used in Metaverse-based environments, financial transactions, and political discourse, facilitating deceptive engagements that are indistinguishable from legitimate interactions \cite{tariq2023deepfake}.
    
\end{enumerate}

\paragraph{Inadequate Multi-factor Authentication} 
To provide a seamless Metaverse experience, it is important to ensure fast, efficient, and trusted cross-platform authentication \cite{dionisio20133d, thakur2024blockchain}. Such cross-platform authentication allows users to roam around among several virtual environments such as Roblox, Fortnite, and Minecraft \cite{oh2023secure, yang2023secure}. Moreover, it allows users to exchange digital assets and transfer avatars among the Metaverse platform. However, most of the current systems lack corss-platform mutli-factor authentication \cite{yang2023secure}.  Inadequate multi-factor authentication (MFA) in the Metaverse poses a significant risk to user information. Poor MFA could allow attackers to steal user identities using stolen or weak credentials. 

\paragraph{Privacy Threats}
Privacy threats in the Metaverse include extensive and intrusive data collection of sensitive information such as biometrics, movements, and physical surroundings.  This sensitive information can be used for surveillance, profiling, and identity theft. 

\begin{enumerate}
    \renewcommand{\labelenumi}{\roman{enumi}.}

    \item \myparait{Pervasive Data Collection:}
    Immersive interaction with avatars in the Metaverse requires detailed and continuous user profiling \cite{falchuk2018social}. Data such as facial expressions, eye and hand movements, speech patterns, and biometric data are used to capture information at an extremely granular level. Human-computer interaction (HCI) and extended reality (XR) technologies allow for real-time user tracking by analyzing bodily movements and personal characteristics \cite{shang2020arspy}. For instance, the Oculus headset can accurately track head orientation, map physical surroundings, and monitor user position with great precision. 
    However, the vast amounts of sensitive data collected by these headsets could be exploited for serious crimes, such as identity theft, surveillance, and behavioral manipulation.

    \item \myparait{ Evolving Data Types:}
    The Metaverse generates various types of data from multiple sensors and devices. For instance, eye movement, facial expression, and head movement data are collected to provide a full immersive user experience \cite{kumar2008second, cheng2024user}. Moreover, devices such as VR headsets and haptic gloves are capable of capturing sensitive biometric data, including iris patterns, fingerprints, and other personal identifiers. The capability of these devices poses significant challenges in collecting, managing, and storing large amounts of sensitive user data \cite{cheng2024user}. The security of these devices themselves becomes challenging, since any compromise might result in data breaches, unauthorized access, or even cause harm to users. Therefore, ensuring secure device design and robust data protection mechanisms is essential to safeguarding user trust and safety in immersive environments. 
    
    \item \myparait{Threats to Personal Safety:}
    The combination of physical devices and immersive technologies in the Metaverse introduces new vulnerabilities that could endanger users' safety and well-being. Hackers can track users' daily activities and real-time whereabouts by taking advantage of wearable technology, XR headsets, and indoor sensors \cite{casey2019immersive}. This might help increase in physical crimes such as stalking or burglaries. According to a worrying scenario provided by the XR Security Initiative (XRSI), enemies might change VR hardware settings, including resetting the physical bounds of a device \cite{initiative2020xrsi}. This manipulation could put users' physical safety at serious risk by causing them to inadvertently approach hazardous real-world areas, such as busy streets or staircases. 


\end{enumerate}


\paragraph{Social Engineering}
Metaverse platforms rely heavily on continuous data collection, including location, environment sensing, communication logs, and interaction histories. This data may be intercepted, disclosed, or misused if appropriate security measures are not undertaken. Moreover, attackers may launch social engineering and content manipulation attacks to gain access to user information. Commonly occurring attacks are data tampering,   phishing, and manipulations \cite{huang2023security}. 

\begin{enumerate}
    \renewcommand{\labelenumi}{\roman{enumi}.}
    \item \myparait{Phishing: }
    In the Metaverse, attackers take advantage of immersive environments, avatars, and interactive content to perform phishing attacks. Phishing in the metaverse extends beyond traditional email or website scams. Attackers may pretend to be virtual service providers, friends, or administrators to trick users into disclosing sensitive information \cite{manoharan2025metahuman}.  In contrast to conventional phishing, metaverse phishing can take advantage of behavioral and visual clues \cite{manoharan2025metahuman}. For example, a malicious avatar can guide a user to a fake marketplace or display deceptive pop-ups in XR devices. Such attacks are especially risky because immersive surroundings boost trust and decrease suspicion, making users more susceptible to manipulation.
    
    \item \myparait{Psychological Manipulation: }
    Psychological manipulation takes advantage of the immersive, interactive, and highly engaging features of the metaverse to take advantage of the feelings and cognitive processes of a user \cite{ali2023metaverse}. 
    Attackers can disseminate false information, instill panic, or incite destructive behavior by using convincing designs, fake social proof, or AI-generated avatars. For instance, hostile actors may employ simulated settings or realistic deepfakes to propagate misleading information or even radicalize individuals.     Furthermore, compared to traditional media, extended exposure to manipulative content in immersive settings may intensify its effects \cite{huang2023security}.  In addition to compromising user security and privacy, this threat poses larger moral and cultural issues, such as encouraging extremism or financial exploitation.

    \item \myparait{Data Tampering Attack:} 
    Data is exchanged across the physical, virtual, and cyber components of the Metaverse. During the exchange of data in the Metaverse or sub-Metaverse, an attacker can modify data. Attackers may attempt to alter, forge, replace, or delete raw data throughout the entire lifecycle of Metaverse services. Such alteration disrupts the normal operations of users, avatars, or connected physical entities \cite{su2020lvbs}. Moreover, attackers may evade detection by manipulating log files or falsifying message-digest outputs, effectively concealing their malicious activities and erasing traces of their presence within the virtual environment \cite{su2020lvbs}. As a result, integrity mechanisms are essential for verifying and detecting any unauthorized modifications to data.

\end{enumerate}

\subsubsection{Avatar}
An avatar is a digital representation of a user in the Metaverse. It serves as a virtual embodiment for interaction, social connection, and self-expression within immersive digital environments. Users can customize their avatars to be realistic, fantastical, or an idealized version of themselves.

\paragraph{Avatar Authentication} 
Verifying avatar in the Metaverse is challenging compared to traditional authentication systems \cite{falchuk2018social}.  Avatar verification involves matching the facial features, voice, and video footage, and several other biometric features \cite{yang2023secure, tariq2023deepfake}. Recent developments in AI models helps attackers to create chatbots capable of mimicking a real avatar \cite{zhang2024anti}. These realistic chatbots can also imitate the appearance, voice, and behaviors of a real avatar in virtual platforms such as Minecraft and Roblox. As a result, many avatar authentication systems rely on personal information for identity verification. However, this approach introduces additional security concerns, as the misuse or compromise of such sensitive data can lead to identity theft and unauthorized access across platforms.

\paragraph{Unauthorized access to Virtual Spaces} 
Metaverse platforms are expected to generate several types of biometric data, daily routines, and user behavior patterns for personal profiling \cite{xu2022wireless}. Virtual Service Providers (VSPs) across different sub-metaverses access real-time profiling data of users to provide seamless and personalized experiences \cite{yu2018leveraging}. However, such user profiling also raises serious privacy and security concerns. Malicious VSPs may attempt to gain unauthorized access to the profiling information for personal gain. For example, attackers could exploit vulnerabilities such as buffer overflow attacks to unlawfully access profiling data.

\paragraph{Malicious Virtual Assets}
To minimize resource consumption, selfish users or avatars in Metaverse may contribute to low-quality User-Generated Content (UGC), which degrades the overall user experience \cite{hussain2023impact}. Such behavior of the users/avatar may result in unrealistic or inconsistent representations within the virtual environment.
For instance, these users may provide inaccurate or highly non-independent and identically distributed (non-IID) data during the collaborative training of content recommendation models, which results in incorrect and unreliable content generation \cite{oh2024enhances}. 


\paragraph{Threats to Virtual Trading}
Service trust in user-generated content (UGC) and virtual object trading is a significant concern in the open virtual marketplace, as avatars can often interact without prior trust or transaction history \cite{liao2021digital}. Repudiation and refusal to pay are two fraud risks that are intrinsic to this ecosystem, especially when users and stakeholders are interacting. Maintaining the authenticity and trustworthiness of digital assets is essential to ensure service reliability.  Malicious users might, for instance, buy user-generated content (UGC) or virtual goods on websites like Minecraft. Then unlawfully copy and resell them for a profit, infringing on ownership rights and compromising the integrity of the marketplace \cite{wanick2023brand}. 

\paragraph{Threats to Digital Asset Ownership}
The decentralized nature of the Metaverse presents significant challenges to the generation, pricing, secure trading, and traceability of digital assets. The lack of a central authority further increases the complexity of the enforcement of ownership rights and dispute resolution. Non-Fungible Tokens (NFTs) provide a potential solution, as they offer distinct, irreplaceable, and tamper-resistant identities that guarantee the provenance and authenticity of assets \cite{wang2021non}. 
Despite these advantages, NFTs are not immune to security threats; they are frequently suffering ransomware, scams, and phishing attacks. 
Adversaries may exploit the lack of cross-chain governance by minting identical NFTs on multiple blockchains simultaneously, undermining their uniqueness and value \cite{popescu2021non}. Additionally, malicious actors may engage in value manipulation schemes and practices that resemble decentralized finance scams. These risks underscore the urgent need for enhanced validation mechanisms, better marketplace transparency, and improved legal frameworks.

\paragraph{Threats to Digital Economy}
In the creator economy of the Metaverse, well-designed incentive systems are essential for promoting user involvement, stimulating creativity, and enabling equitable resource distribution and digital asset trading. However, players who take advantage of the system to obtain unfair advantages frequently jeopardize the integrity of these incentives, jeopardizing long-term sustainability and economic fairness. Three primary adversarial behaviors are of particular concern. First, in order to obtain disproportionate profits, strategic users or avatars might manipulate supply and demand dynamics in digital markets \cite{zhang2021privacy}. 

\subsection{Technology}
Several security risks in the Metaverse arise from vulnerabilities in the hardware, software, and interaction mechanisms that enable immersive experiences. This category contains vulnerabilities in Metaverse programs, APIs, and rendering engines, as well as attacks aimed at XR headsets, IoT devices, haptic systems, and related peripherals. Attackers can obtain unauthorized access, run malicious code, or disrupt human interactions by manipulating sensors, taking advantage of insecure software setups, or taking advantage of firmware defects \cite{meta-privacy-survey:wang2022survey}. Furthermore, vulnerabilities in platform interfaces and smart contracts might act as gateways for remote exploitation, data manipulation, or privilege escalation, endangering linked real-world systems and virtual environments.

\subsubsection{Extended Reality (XR): VR, AR, and MR}
Extended Reality (XR) combines virtual reality (VR), augmented reality (AR), and mixed reality (MR). XR blends physical and virtual objects to provide an immersive experience to users. Despite all these benefits of XR, they are susceptible to many attacks. 
 

\paragraph{XR Firmware Exploits}
Firmware in XR headsets and other devices is often a prime target because it operates close to the hardware and typically receives infrequent security updates. Attackers can bypass advanced security measures, install persistent malware, or completely take over the device by taking advantage of firmware flaws \cite{zhang2025industrial}.  In IoT settings, this might allow remote activation or data exfiltration without user awareness. In XR systems, it could allow malicious code to intercept sensory data or change displayed content \cite{jaber2022security}. 
Limited computational resources and proprietary firmware designs further complicate patching and make such devices more susceptible to long-term compromise.

\paragraph{Biometric Data Leaks}
In the Metaverse, immersive experiences are mostly based on sensors and gadgets that capture distinct physiological and behavioral characteristics of users \cite{huang2023security}. Biometric data leaks pose a serious privacy and security concern in the Metaverse. These consist of voiceprints, iris scans, facial recognition patterns, locomotion, and even neurological signals from brain-computer connections. Biometric data cannot be easily revoked or changed once compromised, making its leakage particularly severe \cite{huang2023security}. When attackers obtain such data, they can commit identity fraud, sell private information on illegal markets, or mimic users across platforms. 
Moreover, leaked biometric information may enable advanced profiling and surveillance, exposing users to long-term risks that extend beyond the virtual world.

\subsubsection{Communication Networks}
Metaverse is vulnerable to traditional communication network threats due to its foundational reliance on the existing Internet and wireless technologies. As Metaverse applications are data-intensive and latency-sensitive, this makes the underlying infrastructure more vulnerable to sophisticated attack vectors. For instance, adversaries may be able to intercept private user information such as location, identity, and behavior by eavesdropping on unprotected wireless connections. These threats underscore the importance of incorporating advanced security protocols to safeguard the communication backbone of Metaverse platforms.

\paragraph{SPoF Attacks}
Centralized architectures provide a number of benefits in the development of Metaverse systems, including simplified management of users and avatars, faster data processing, and more economical operations. However, there are substantial security and resilience trade-offs associated with these advantages. Centralized systems are vulnerable to Single Points of Failure (SPoF) \cite{wang2021blockchain}. Millions of users could be impacted if a root server is physically damaged or taken offline, disrupting the entire service. Furthermore, centralized control poses significant questions about transparency and trust. Especially when it comes to overseeing the transfer of digital assets, currencies, and virtual commodities between various and compatible virtual worlds \cite{noveck2011single}. Users must rely on the integrity and fairness of the central authority in the absence of a trustless infrastructure. This centralized model limits the vision of a decentralized and user-driven Metaverse. In order to develop safe, scalable, and trust-free interactions in virtual ecosystems, more robust, distributed architectures are necessary. 

\paragraph{DDoS Attacks}
Metaverse comprises a vast number of small, wearable devices; these end-devices are attractive targets for attackers aiming to launch large-scale cyberattacks \cite{bertino2017botnets}.  Attackers may utilize these devices to initiate Distributed Denial-of-Service (DDoS) attacks. These attacks could potentially cause significant network outages and service interruptions by overloading centralized servers with traffic, temporarily making Metaverse environments unavailable.
The centralized nature of many Metaverse systems results in a lack of the distributed resilience required to tolerate large traffic spikes, making this vulnerability even worse \cite{kuo2023metaverse}. Furthermore, due to blockchain technology having limited storage capacity and transmission bandwidth, some Non-Fungible Token (NFT) capabilities are offloaded to off-chain platforms in real-world applications. This off-chain execution increases efficiency; however, it brings additional vulnerabilities. Using DDoS attacks, adversaries can disrupt vital services related to NFT transactions, ownership verification, or content access. These scenarios underscore the need for decentralized architectures, robust edge-device security, and hybrid blockchain solutions to maintain both scalability and resilience against evolving threats in the Metaverse.

\paragraph{Sybil Attacks}
Sybil attacks pose a serious threat to the integrity and effectiveness of Metaverse services by allowing attackers to create or exploit multiple fake or stolen identities to gain an unfair advantage \cite{cao2024eden}. Such malevolent actors have the ability to interfere with voting-based governance procedures on digital platforms, affect blockchain consensus mechanisms, and manipulate reputation systems. Sybil adversaries can distort collective decision-making processes by controlling a large number of fraudulent identities. Such attacks reduce trust among genuine users and potentially seize control of critical components of the Metaverse infrastructure \cite{kim2022novel}. For example, in blockchain-based systems integrated into the Metaverse, adversaries could generate a sufficient number of Sybil nodes to outvote legitimate participants. Such attacks can halt the flow of transactions, disrupt decentralized services, and undermine the reliability of digital governance. As a result, Sybil attacks not only compromise fairness and transparency but also threaten the scalability and sustainability of decentralized systems in the Metaverse. Addressing these threats requires the implementation of robust identity verification, trust mechanisms, and reputation management systems that are resistant to identity forgery and duplication.

\paragraph{Man-in-the-middle (MitM) Attacks and Eavesdropping}
MitM attacks and eavesdropping in the Metaverse occur when an adversary intercepts or manipulates data transmissions between users, devices, or servers \cite{su2021secure}. 
Even a brief interception can reveal private passwords, biometric information, or in-world communications in latency-sensitive settings such as cloud-rendered scenes or XR streaming \cite{vondravcek2023rise}. In order to obtain unencrypted data or introduce malicious content, attackers may breach edge nodes, unprotected Wi-Fi, or susceptible network paths. This not only threatens user privacy but can also alter virtual interactions. 

\paragraph{Jamming Attacks}
5G and emerging 6G networks provide the high bandwidth and low latency required for immersive Metaverse experiences. However, they are vulnerable to jamming attacks due to their wireless nature \cite{yang2023secure}. 
Attackers can interfere with communication channels by sending out powerful radio signals in the same frequency ranges, which can result in poor performance, disconnections, or total denial of service \cite{ali2023metaverse}. These attacks have the potential to disrupt cloud gaming sessions, IoT device coordination, and XR streaming for a large number of users at once \cite{wang2021blockchain}. 
Metaverse often relies on continuous connectivity, and jamming can be used as a tool for large-scale disruption.

\paragraph{SDN Controller Attacks}
Software-Defined Networking (SDN) allows for flexible and programmable network administration for Metaverse infrastructure by separating network control from data transmission. However, the SDN controller can become a high-value target because it is the primary decision-making component \cite{wang2021blockchain}. 
Compromising the controller allows attackers to manipulate traffic flows, reroute data to malicious endpoints, or entirely disable security policies. Such attacks may result in sensitive data interception or widespread denial-of-service attacks. Since SDN frequently serves as the foundation for cloud and edge computing in the Metaverse, a breach of an SDN controller may have repercussions that affect numerous services and geographical areas.
 
\paragraph{Network Slice Isolation Failures}
In 5G/6G, network slicing enables operators to design separate, virtualized network segments for particular Metaverse applications, such as IoT sensor networks or XR streaming \cite{de2023survey}. 
However, vulnerability in slice configuration, isolation, or access control can allow one compromised slice to impact others. 
For instance, a vulnerability in a low-priority IoT slice could be exploited to gain access to a high-priority XR slice \cite{de2023survey}. Additionally, improperly designed slicing may result in illegal cross-slice data access, resource starvation, or a decline in service quality. Misconfigured slicing can also lead to resource starvation, degraded service quality, or unauthorized cross-slice data access. 
These vulnerabilities compromise the reliability and security promises of network slicing, posing risks to the Metaverse.

\subsubsection{Digital Twins (DT)}
A digital twin is a virtual representation of a real-world physical object, system, or process \cite{wu2021digital}. It builds a dynamic, changing model that replicates the real one using real-time input from sensors and other sources. 

\paragraph{Unauthorized Access to DT Models}
Confidentiality and intellectual property are seriously threatened by unauthorized access to digital twin (DT) models in the Metaverse \cite{yao2022metaverse}. DTs replicate real-world assets, processes, or infrastructures. Attackers obtaining unauthorized entry can steal sensitive design data, manipulate proprietary models, or misuse operational parameters. A compromised DT can be used to simulate attacks or disrupt critical infrastructure operations \cite{yao2022metaverse}. This risk is increased by inadequate access controls, weak authentication procedures, and unsafe data sharing protocols, which make DT models appealing targets for bad actors.

\paragraph{Manipulation of DT Simulations}
DT simulation manipulation can produce erroneous results that mislead Metaverse decision-making. To create misleading findings, attackers can change simulation parameters, insert malicious code, or skew the synchronization between digital and physical objects \cite{kuru2023metaomnicity,  liao2021digital}. 
For example, in smart manufacturing or intelligent transportation, manipulated simulations could falsely indicate safe operating conditions, resulting in severe physical-world consequences. Such attacks undermine trust in DT-driven forecasts, planning, and control systems in a variety of industries by posing a danger to integrity and dependability.

\subsubsection{Artificial Intelligence (AI)}
Artificial Intelligence (AI) plays a pivotal role in shaping the metaverse by enabling intelligent interactions, dynamic content creation, and adaptive personalization. AI techniques such as natural language processing (NLP), computer vision, and reinforcement learning allow avatars to engage in realistic conversations, recognize user gestures, and learn from behavioral patterns, thereby creating immersive and human-like experiences \cite{adversarial}.

\paragraph{Data Injection Attack}
The injection of falsified information into the original data can mislead the Metaverse systems. For instance, AI-assisted content generation enhances immersion in the Metaverse platform but also opens new opportunities for adversarial manipulation \cite{liang20162015, siddiqi2024multichain}. Attackers may push adversarial training samples or poisoned gradients during centralized or distributed AI training processes, leading to the creation of biased or malfunctioning models. These biased models can produce incorrect feedback or instructions, posing risks to both virtual and physical components. In extreme cases, such false training and inference of AI models may endanger user safety. Excessive voltage could be transmitted to wearable XR devices, potentially causing physical harm.

\paragraph{Malicious Autonomous Agents}
In the Metaverse, malicious bots and compromised non-player characters (NPCs) can be weaponized to manipulate interactions, spread misinformation, or deceive users into disclosing sensitive information \cite{han2022cheating}. In contrast to conventional gaming NPCs, Metaverse NPCs could be AI-powered and social interaction-focused, making it challenging for users to differentiate between benign and malevolent entities \cite{han2022cheating}. Additionally, these bots can replicate human behavior, enhancing phishing attempts, disinformation operations, and even financial crime in virtual marketplaces.

\subsubsection{Blockchain}
Blockchain plays an important role in enabling trust, transparency, and decentralization within the Metaverse platforms. It supports secure digital asset ownership, a secure virtual economy, and decentralized governance. However, blockchain integration also introduces security and privacy challenges. 

\paragraph{Smart Contract Exploitation}
Blockchain-powered Smart contracts are frequently used by metaverse platforms to control virtual assets, transactions, and governance. Contract logic flaws, such as reentrancy vulnerabilities or careless input handling, can be used to steal money, change ownership records, or get around specified limitations \cite{brown2018security, alkhalifah2021mechanism}. In the same manner, apps with inadequate role-based access control may enable attackers to escalate privileges and get administrative powers without permission \cite{sun2024gptscan}. As blockchain transactions are immutable, an attacker might alter crucial platform settings, manipulate in-game economics, or interfere with vital services after such escalation occurs with little to no evidence \cite{seo2024space}.

\paragraph{Private Key Theft}
The private key serves as the sole authentication mechanism for transactions in the Metaverse. The security of blockchain-based assets and identities in the metaverse depends heavily on the protection of private keys \cite{ryu2022design}.
An attacker can use phishing to steal the private key of a user and gain complete control over that user’s assets and digital identity. Private key theft results in an irreversible loss of ownership, in contrast to older systems where compromised credentials can frequently be revoked or reset \cite{din2024securing}.  This makes the Metaverse a high-value target for attackers and calls for more robust key management and authentication systems.

\paragraph{Fraudulent Digital Assets}
In the metaverse, non-fungible tokens (NFTs) facilitate the ownership and exchange of virtual property, art, and digital goods \cite{wang2021non}. 
However, the rapid growth of NFTs has led to widespread fraud, including the minting of counterfeit assets, unauthorized tokenization of copyrighted works, and rug-pull schemes \cite{popescu2021non}. 
Fraudulent NFTs not only undermine user trust but also cause instability in virtual economies by inflating asset bubbles and enabling illegal trade \cite{wang2021non}. 
The absence of cross-platform interoperability and standardized verification mechanisms makes it challenging to guarantee the provenance and validity of digital assets across metaverse ecosystems.

\subsubsection{Internet of Things (IoT)}
IoT plays a crucial role in the Metaverse by enabling real-world data to seamlessly integrate into virtual environments. IoT devices, equipped with sensors and software, collect data from physical spaces and transmit it to the Metaverse, enhancing the user experience with real-time information and interactions \cite{vu2020learning}. 

\paragraph{Compromised Devices}
In the Metaverse, wearable devices allow avatars to mimic real-world behaviors such as making natural eye contact, recognizing hand gestures, and reflecting facial expressions in real time \cite{shang2020arspy}. Although this improves immersion, there are serious privacy and security concerns as well. These devices capture highly sensitive information that can reveal the behaviors of the user. Compromised wearable devices, such as VR headsets or haptic gloves, further enhance the attack surfaces. These rogue devices can serve as gateways for data breaches and malware attacks. Avatars may unintentionally turn into tools for unauthorized data collection when influenced by compromised hardware, gravely violating user privacy. For instance, sophisticated gadgets like haptic gloves and Oculus headsets can detect subtle movements such as finger gestures and eye movements \cite{blow2020study}. This data could be used by attackers to reconstruct user actions, including decoding private data such as PINs or passwords typed into virtual keypads \cite{blow2020study}. Such capabilities underscore the need for secure hardware standards, continuous device authentication, and stringent monitoring protocols.

\paragraph{Sensor Spoofing Attacks}
Sensor spoofing involves feeding false data into the sensors to manipulate their behavior or the environment they present. Spoofing attacks in the Metaverse could use environmental cameras, motion sensors, GPS modules, or eye trackers to alter spatial placement, insert fictitious environmental data, or mimic user gestures \cite{ali2023metaverse}. In mixed-reality environments, such manipulation may lead to bodily injury, misdirect autonomous systems, or produce deceptive or malicious experiences \cite{chen2023metaverse}. Sensors frequently trust incoming data without thorough verification, making these attacks hard to detect. They can be carried either wirelessly or through interference from the environment.

\paragraph{Insecure IoT Interfaces}
APIs are essential for data sharing, asset management, and service integration in Metaverse platforms and applications. APIs with inadequate security (i.e., missing input validation, authentication, or encryption) are susceptible to illegal access, data leaks, and virtual environment manipulation \cite{ali2023metaverse}. 
Attackers may be able to increase access, alter user assets, or interfere with platform operations by taking advantage of application vulnerabilities such as injection errors or improper permissions \cite{huang2023security}.
Furthermore, a single unsecured endpoint can serve as a focal point for more extensive breaches because many Metaverse APIs communicate with several services and outside technologies.

\subsubsection{Cloud \& Ubiquitous Computing}
Cloud and ubiquitous computing form the backbone of the Metaverse by providing large-scale storage, computing, and networking resources. By distributing resources closer to users, they enhance scalability, responsiveness, and accessibility, making the Metaverse a persistent and interconnected digital ecosystem.

\paragraph{Data Leakage}
Metaverse relies on the collection and processing of enormous volumes of data to create and render realistic avatars and immersive virtual worlds. These data are generally collected from users and their surrounding environments. However, this data-intensive process involves significant privacy risks, as it involves the processing and handling of highly sensitive personal information \cite{li2021verifiable}. To train personalized avatar models, users' private data is aggregated from multiple sources into centralized storage systems. Such central processing may lead to privacy breaches and potentially violate data protection laws \cite{gdprinfoGeneralData}. Although identifiable information is removed, attackers can still analyze the processed outputs, such as generated avatars or user behavior patterns. 
These inference attacks underscore the need for stronger privacy-preserving mechanisms. 

The storage of private and sensitive information, such as user profiling data, on cloud servers or edge devices, poses significant privacy risks. These storage locations often store data from millions of users, which makes them vulnerable to cyberattacks \cite{wei2020ldp}. Attackers may push frequent queries and utilize advanced attack techniques to deduce private user information, even without direct access to raw data. Furthermore, attackers can launch distributed denial-of-service (DDoS) attacks exploiting vulnerabilities in cloud or edge infrastructure \cite{bertino2017botnets}. Such attacks compromise the availability and integrity of stored information. In 2006, the customer database Second Life was compromised, revealing unencrypted usernames and physical addresses, along with encrypted payment information and passwords \cite{ludlow2007second}. The incident highlights the real-world implications of insufficient data protection. 

These security risks emphasize the need for robust data encryption, access control, and anomaly detection mechanisms in the Metaverse.
In Metaverse systems, a substantial amount of personally identifiable information (PII) is collected through wearable devices such as head-mounted displays (HMDs). Collected data are transmitted via both wired and wireless communication channels. Maintaining the privacy of this data is essential to preventing sensitive user information from being accessed by unauthorized people or services \cite{ometov2016facilitating}. Even while encryption mechanisms are usually used to safeguard data while it is being transmitted, weaknesses can still be exploited by attackers. For instance, skilled attackers can eavesdrop on the transmission channels to intercept raw data \cite{wei2020ldp}. They can also employ sophisticated attacks to identify user locations and behavioral patterns \cite{wasserkrug2008inference}. These risks highlight the importance of developing robust and privacy-preserving communication frameworks. The development of advanced defense mechanisms can help safeguard the user's information against evolving threats in immersive virtual environments.

\paragraph{Insider Threats}
Malicious insiders, including administrators or staff members with privileged access, can compromise Metaverse cloud systems. Unauthorized data exposure, virtual asset manipulation, or deliberate service sabotage are all possible outcomes of insider threats \cite{hu2015dynamic}. Insider attacks are more difficult to identify and can go unnoticed for extended periods of time because they frequently evade traditional perimeter defences, endangering Metaverse security and governance \cite{he2022survey}.


\subsection{Governance Challenges}
Avatar interactions in the Metaverse, including content production, social interaction, and involvement in the virtual economy, must follow established digital norms and regulatory frameworks \cite{almeida2021ecosystem}. Such norms and regulatory frameworks ensure moral and legal behavior, much like social norms and regulations govern behavior in the real world. Effective governance is essential to maintain a safe and orderly virtual environment. However, several risks that could jeopardize system security and efficiency may surface during the oversight and regulatory procedures. These include weak enforcement mechanisms, jurisdictional ambiguity, regulatory capture, and technological loopholes that can be exploited by malicious actors. Without robust digital governance, the Metaverse could become a space where harmful behaviors, fraudulent transactions, and legal violations occur with impunity \cite{almeida2021ecosystem}.

\subsubsection{Lack of Legal Frameworks for Virtual Crimes}
One of the major challenges in governing the Metaverse lies in the difficulty of determining whether a virtual crime should be treated the same as its real-world counterpart. This ambiguity makes it more difficult to use current legal frameworks directly to control and penalize criminal behavior in virtual environments \cite{hendaoui20083d}. For example, actions such as the use of abusive language by an avatar can be interpreted as verbal abuse, regardless of whether they occur in the physical or virtual world \cite{qin2025identity}. However, more complex cases, such as virtual stalking or harassment, blur the boundaries between real and virtual misconduct. Such behaviors may differ greatly from real-life circumstances in terms of their nature, impact, and context in immersive settings. This further raises concerns regarding how these behaviors should be characterized and dealt with legally. As a result, it might be necessary to modify current laws and create new legal standards for the Metaverse.  For lawmakers and regulators, this evolving digital landscape calls for a reevaluation of traditional legal principles to ensure justice, accountability, and user protection in the Metaverse.

\subsubsection{Regulatory Misconduct}
In the Metaverse, regulators themselves can become a source of dysfunction if they act dishonestly or irresponsibly, potentially leading to system paralysis. Clear reward and punishment systems should be put in place, and their authority must be monitored, to stop such dishonest behaviors. To maintain fairness and transparency, the Metaverse systems should democratically and decentralizedly enforce these norms. A potential remedy is provided by smart contracts, which automate regulations without the need for reliable third parties \cite{bai2021public}.  However, they introduce new risks, including information leaks, mishandled exceptions, and vulnerabilities such as short address attacks and reentrancy attacks \cite{sayeed2020smart}. Despite their potential, smart contracts need to be properly designed and routinely audited to guarantee safe and trustworthy Metaverse governance.

\subsubsection{Policy Inconsistency}
To prevent the concentration of regulatory power, collaborative governance models are considered more ideal for managing large-scale Metaverse ecosystems \cite{huang2019software}.  These methods promote fairness and transparency by dividing up decision-making among several entities. But even in these kinds of systems, regulator collusion still remains a serious risk.  For example, a group of dishonest regulators may plan to use a wormhole attack to isolate or disable another regulator, thereby undermining the integrity of the system and interfering with governance functions. This emphasizes that even in frameworks for collaborative governance, strong safeguards and trust measures are necessary.

\subsubsection{ Digital Forensics Limitations}
Digital forensics in the Metaverse refers to the process of virtually reconstructing cybercrimes by identifying, gathering, and analyzing evidence from both real and virtual environments \cite{li2021toward}. Significant difficulties are presented by the high levels of dynamism, anonymity, and interoperability among many virtual worlds. Diverse behavior patterns and limited traceability of users and avatars make tasks such as tracing actions, confirming identities, and attributing behaviors to specific entities more difficult. Furthermore, the blurred boundaries between real and virtual worlds further complicate investigations, making it difficult to distinguish fact from fabrication. 
For instance, malicious actors can exploit AI technologies to generate fake news, voices, or videos to deceive the public and spread misinformation \cite{tariq2023deepfake}.

\subsubsection{Infrastructure Safety Risks}
In the highly integrated architecture of the Metaverse, hackers can exploit software or system vulnerabilities to gain unauthorized access through compromised devices such as XR headsets, sensors, or connected smart systems \cite{hu2015dynamic}. These compromised devices may be used as gateways for more complex assaults, such as Advanced Persistent Threats (APTs), which are focused, protracted operations frequently directed at vulnerable systems \cite{vellaithurai2014cpindex}. Such breaches could allow adversaries to gain access to vital national infrastructures, like high-speed rail networks or power grids, endangering vital services, national security, and public safety. This highlights the urgent need for securing all layers of the Metaverse ecosystem. 

\subsubsection{Intellectual Property Violations}
The definition of intellectual property (IP) differs in the Metaverse from the physical world. The concept of IP in the Metaverse must be redefined to handle its evolving nature and increasing scale. To handle user-generated content (UGCs) and AI-generated content (AIGCs), an IP framework tailored to the unique characteristics of virtual environments is required \cite{hendaoui20083d}. One of the major challenges lies in the dissolution of geographic boundaries, which complicates jurisdiction and enforcement.
As the Metaverse expands across national borders, establishing ownership and legal protection across different legal systems becomes increasingly difficult \cite{hendaoui20083d}. For example, Concerns about identity rights and unapproved commercial exploitation have already surfaced in fights over the use of celebrity likenesses in video games \cite{btljRightPublicity}.
As economic value tied to digital identities and assets is growing, such conflicts are expected to increase dramatically in the future Metaverse. A globally coordinated legal and policy framework is needed to address IP challenges in virtual ecosystems.

\subsubsection{Adverse Social Impacts}
The Metaverse presents an exciting vision of a digitally enhanced society, but it also raises important ethical and social issues that could have a detrimental effect on human life. The immersive and engaging nature of the Metaverse increases the risk of user addiction, as individuals may find it difficult to detach from persistent virtual environments. Furthermore, the anonymity and wide reach of these platforms exacerbate problems such as digital extortion \cite{valluripally2021modeling}, child exploitation, cyberbullying, cyberstalking \cite{leenes2007privacy}, and rumor propagation \cite{zhu2020activity}. The immersive realism and lack of adequate regulation further complicate the ability to detect and prevent these harmful behaviors. The possibility of extremist and terrorist groups using the Metaverse is one particularly concerning issue \cite{weimann2023metaverse}. At a low cost and with a lower chance of discovery, the immersive virtual worlds open up new possibilities for recruitment, training, coordination, and planning. For instance, terrorists might carefully plan attacks and determine escape routes by using computerized models of actual structures for simulated training. This degree of virtual practice can greatly improve the accuracy and efficacy of real-world terrorist and illegal operations.
Furthermore, the Metaverse presents difficult ethical issues as it grows more independent and may be controlled by AI-driven algorithms \cite{leenes2007privacy}. As seen in the science fiction movie "The Matrix\footnote{https://en.wikipedia.org/wiki/The\_Matrix}", when AI systems govern interactions and norms, problems such as racial and gender bias, algorithmic discrimination, and a lack of human oversight may arise.  These outcomes could lead to systemic injustices in digital societies, reinforcing existing inequalities or even creating new forms of digital oppression. Proactive governance, open AI systems, and inclusive ethical frameworks are necessary to mitigate these hazards.

\begin{tcolorbox}[boxrule=0.5pt,boxsep=1pt,left=2pt,right=2pt,top=2pt,bottom=2pt]
\noindent\textbf{RQ2 Summary:}
Compared to conventional digital platforms, the Metaverse offers a much larger attack surface.  The immersive, decentralized, and permanent characteristics provide new vulnerabilities in the levels of identity, data, infrastructure, and interaction.  The emergence of unregulated dark virtual environments, biometric data leaks, authentication errors, deepfake manipulation, avatar impersonation, and economic exploitation is one of the main threats.  Furthermore, in these real-time scenarios, conventional security measures are either inadequate or incompatible. Tables \ref{tab:metaverse-cia-threats-user}, \ref{tab:metaverse-cia-threats-technology}, and \ref{tab:metaverse-cia-threats-governance} summarize the existing security threats and vulnerabilities in Metaverse. 

\end{tcolorbox}

\section{Security Mechanism in Metaverse (RQ3)}  \label{sec:rq4}
Protecting the Metaverse's data, infrastructure, and user interactions has become crucial as it develops into a persistent and immersive digital world.  Due to its multidimensional character, which combines decentralized platforms, blockchain, IoT, extended reality (XR), and artificial intelligence (AI), the Metaverse is vulnerable to a wide range of threats.  In response, defense methods and technologies have been devised and put into practice by researchers and developers. These range from identity verification schemes and cryptographic protocols to AI-driven detection systems and privacy-preserving frameworks.  The goals of these solutions are to provide trust, privacy, integrity, and authentication across the Metaverse platforms.

\subsection{Countermeasures to User Threats}
In the Metaverse, secure identity management is very important for user/avatar interactions and service provisioning. Three of the most widely used identity types are Centralized Identity (e.g., Gmail), Federated Identity \cite{jensen2012federated}, and Self-Sovereign Identity (SSI) \cite{samir2021dt}. Federated Identity supports cross-domain access and reduces repeated data input, and SSI, giving users full control over their digital identity and consent-based data sharing.

\subsubsection{Blockchain-Based Access Control}
Blockchain plays a pivotal role in enhancing identity security and access control in the Metaverse. By leveraging decentralized identifiers (DIDs) and smart contracts, blockchain enables secure, verifiable, and tamper-proof identity systems. For instance, the Metaverse-AKA protocol introduced by Yao et al. \cite{yao2022metaverse} uses elliptic curve cryptography (ECC), verifiable credentials (VCs), and decentralized key management to enable seamless, privacy-preserving authentication across multiple Metaverse domains. This cryptographic protocol defends against impersonation, replay, and man-in-the-middle (MITM) attacks.
Blockchain is essential for improving access control and identity security in the Metaverse.  Blockchain technology uses smart contracts and decentralized identities (DIDs) to provide a secure, verifiable, and impenetrable identification system.  For example, the Metaverse-AKA protocol, which was presented by Yao et al. \cite{yao2022metaverse}, enables smooth, privacy-preserving authentication across several Metaverse domains by utilizing elliptic curve cryptography (ECC), verifiable credentials (VCs), and decentralized key management.  This cryptographic protocol helps in the prevention of Man-in-the-Middle (MITM), replay, and impersonation attacks. Furthermore, space-based authentication systems such as that proposed by Seo et al. \cite{seo2024space} authenticate both the user and the virtual environment (space) they access. This two-way trust mechanism uses smart contracts and similarity thresholds to authenticate entry into specific virtual spaces, providing spatial-level access control that goes beyond traditional identity checks.

\subsubsection{Biometric Authentication}
Biometric authentication is gaining popularity over time as the Metaverse continues to develop. Biometric information like eye movements, body position, and facial geometry is frequently used for authentication in the Metaverse due to the dependence on avatars and XR devices.  Cardiac biometrics from photoplethysmography (PPG) sensors in wearables achieve a 90.73\% average continuous authentication accuracy \cite{jan2020lightweight}. However, holding such sensitive information could result in biometric spoofing. To mitigate such risk, advanced biometric fusion systems are proposed by Zhang et al. \cite{zhang2024anti} that combine 3D facial features with body biometrics to create robust, disguise-resistant user profiles. These systems outperform single-modality methods in preventing identity spoofing and enable reliable avatar verification. Furthermore, privacy-aware mutual authentication using Hidden Markov Models (HMM) is proposed to mitigate patient data leakage risks \cite{jan2020lightweight}, with its security validated through Burrows–Abadi–Needham (BAN) logic. Bluetooth-based wearable fingerprinting has also been explored \cite{aksu2018identification}, while Arias et al. \cite{arias2015privacy} demonstrated real-world attacks on devices like Google Nest and Nike+ Fuelband, emphasizing the need to secure update channels and disable debug access.
Multi-factor authentication (MFA) is still a crucial element in practice.  To lower the risk of unwanted access, several platforms now provide password, device token, biometric scan, and QR-based validation combinations.

\subsubsection{Smart Contracts and Multisignature Access Control}
In decentralized Metaverse platforms, smart contracts enable safe and verifiable operations.  Gai et al. \cite{gai2022blockchain} proposed a smart contract-based Blockchain Multisignature Lock (BMSL-UAC) mechanism.  This model enables ubiquitous access control across digital institutions in the Metaverse by requiring multiple authorized signatures before granting access to shared data or spaces. This model ensures accountability, traceability, and defence against replay, spoofing, and tampering attacks.

\subsubsection{Cross-domain Identity Authentication}
Shen et al. proposed a blockchain-based scheme using identity-based encryption (IBE) for cross-domain identity authentication \cite{shen2020blockchain}. Proposed systems use anonymous protocols with identity revocation, supported by off-chain storage. Similarly, Chen et al. introduced XAuth, an optimized blockchain framework integrating zero-knowledge proofs for anonymous and efficient cross-domain authentication \cite{chen2021xauth}.

\subsubsection{Wearable and UGC}
Fine-grained authentication techniques are also needed for wearables and UGC access control. Ometov et al. proposed a privacy-respecting delegation-of-use model \cite{ometov2016facilitating}. For UGMCs, Ma et al. \cite{ma2018scalable} developed a scalable CP-ABE-based approach, while for UGVCs, Yang et al. \cite{yang2016time} presented a time-domain attribute-based mechanism. Zhang et al. designed a secure sharing system with proxy re-encryption and watermarking to combat illicit redistribution \cite{zhang2018you}.

\subsubsection{Privacy-Preserving UGC Sharing and Processing}
To ensure the privacy of user-generated content (UGC) sharing and processing, techniques such as differential privacy (DP), federated learning (FL), secure multiparty computation (SMC), homomorphic encryption (HE), and zero-knowledge proofs (ZKP) are widely used. A graph-based local DP method for trending topic recommendations is presented by Wei et al. \cite{wei2020ldp}. Zhang et al. \cite{zhang2021fedsens} provide privacy-aware health sensing by combining FL with RL and regret minimization. ZKP is used by Guan et al. \cite{guan2020blockmaze} to safeguard blockchain transaction data. In order to protect the privacy of multi-user facial data, Xu et al. \cite{xu2015my} use SMC and SVM to address co-photo privacy. A verified HE-based method for edge data processing is designed by Li et al. \cite{li2021verifiable}. Lastly, Wang et al. \cite{wang2020spds} utilize trusted computing with fair-exchange smart contracts for privacy-preserving off-chain UGC processing.

While most works focus on access control, usage control is often overlooked. Wang et al. presented a blockchain-based DPaaS paradigm, allowing the development of data usage policies, their execution through smart contracts, and transparent audits \cite{wang2020spds}. Yu et al. combined UGC sensitivity and user trustworthiness using deep learning and social grouping to train classifiers for personalized privacy settings \cite{yu2018leveraging}. These efforts work together to improve access control, privacy, and identity interoperability in the dynamic Metaverse context.

\subsubsection{Countermeasures to Social Engineering Threats}
The immersive and engaging nature of the metaverse greatly increases the impact of social engineering attacks. Attackers can persuade users or avatars into revealing private information or engaging in destructive activities by taking advantage of human psychological characteristics. For instance, phishing in virtual spaces, impersonating trusted entities, and conducting scams through fraudulent marketplaces or events are very common in the virtual world \cite{de2023survey}. Deepfakes, AI-generated disinformation, and malicious user-generated content (UGC) are examples of content manipulation risks in the Metaverse. In the following, we present the countermeasure to prevent social engineering and content manipulation attacks in the Metaverse.

\paragraph{AI-driven Content Moderation}
Poor data quality degrades user experience and service quality.  Behavioral realism increases perceived character credibility, as demonstrated by Dickinson et al. \cite{dickinson2021experiencing}. Incentive-based methods such as Stackelberg games \cite{xu2020game} and deep reinforcement learning \cite{su2021secure} encourage high-quality contributions. While Du et al. \cite{du2022optimal} used hidden communications for high-quality access techniques, Han et al. \cite{han2022dynamic} suggested a hierarchical game for maximizing DT synchronization for data availability and synchronization.

Furthermore, Artificial Intelligence (AI) is increasingly used to detect malicious activities in real time. Machine learning (ML) based detection method provides a robust and reliable method for automatic threat detection. Intrusion attempts, abnormal behaviors, deepfake attacks, and synthetic inputs are most commonly detected with the help of ML detection systems. AI-based systems pick up on common patterns of user activity and highlight any differences that might point to exploitation or compromise. For instance, Otoum et al. \cite{otoum2024machine} reviewed various ML models designed for Metaverse security tasks such as malware detection, avatar behavior analysis, and privacy violation prediction. ML-based anomaly detection is also useful in protecting systems from adversarial attacks on edge AI systems, especially in 6G-enabled IoT environments. Li et al. \cite{li2023robust} used adversary detection-deactivation procedures to offer a defence strategy for federated learning setups in the Metaverse in collaborative environments. To maintain model integrity, their two-layer model identifies and deactivates participant nodes that provide fraudulent updates.

\paragraph{Misinformation Filtering} 
In Metaverse environments, AI-generated content (AIGC) enhances realism, but adversarial samples pose significant risks. There have been several countermeasures proposed, including reinforcement \cite{sun2020stealthy}, transfer learning \cite{zheng2020efficient}, adversarial representation \cite{mai2020modality}, and virtual adversarial learning \cite{miyato2018virtual}. To ensure the trustworthiness of digital twins, Gehrmann and Gunnarsson \cite{gehrmann2019digital} proposed a state replication model with defined security architecture requirements. A permissioned blockchain and the DT-DPoS protocol were employed by Liao et al. \cite{liao2021digital} to ensure safe transactions in intelligent transportation systems (ITS). To ensure audio data reliability, Zimmermann and Liang \cite{zimmermann2008spatialized} proposed an area-of-interest-based system, and Jot et al. \cite{jot2021rendering} introduced 6 Degrees of Freedom (6DoF) parametric audio scene programming. Furthermore, GAN-detection algorithms and AI-based content moderation systems have been incorporated by multiple platforms to counteract misinformation and impersonation based on deepfakes.  To identify generated content, these technologies examine facial expressions, voice, and video textures. According to Tariq et al. \cite{tariq2023deepfake}, real-time deepfake detection in virtual workplaces and business encounters is critically needed.

\paragraph{Deepfake Detection Algorithms}
Deepfake detection algorithms are becoming increasingly essential to combating AI-generated impersonations. Use of AI to mimic user behavior threatens the authenticity of interactions in the metaverse. Deepfake detection algorithms can help flag these deep-fake contents. These algorithms analyze inconsistencies in visual, auditory, or behavioral patterns to detect synthetic content such as fake avatars, altered voices, or forged facial expressions \cite{tariq2023deepfake}. 
Advanced detection algorithms utilize deep neural networks, temporal analysis of motion patterns, and multimodal cross-verification to identify artifacts that are imperceptible to humans. However, attackers can constantly improve generative models to get around current defences due to the race between deepfake production and detection. Furthermore, in high-interaction settings like virtual meetings or virtual marketplaces, latency and scalability issues arise due to the resource-intensive nature of real-time deepfake detection \cite{oh2023secure}. Therefore, in order to preserve confidence in digital interactions, future metaverse systems need to have detection pipelines with provenance verification and blockchain-based authenticity monitoring.

\paragraph{Secure Data Sharing} 
Multi-user XR settings require secure data sharing. To ensure secure data sharing in an AR environment, Ruth et al. introduced AR sharing controls using HoloLens \cite{ruth2019secure}. Lee et al. identified three ad fraud threats in WebVR and proposed the AdCube defense to mitigate those threats \cite{lee2021adcube}. These techniques show high effectiveness with minimal cost.

\paragraph{Provenance of UGC} 
Data provenance ensures audit trails, source responsibility, and traceability. 
To identify sensor tampering, Satchidanandan and Kumar suggested dynamic watermarking \cite{satchidanandan2016dynamic}. ProvChain is a blockchain-based cloud provenance system proposed by Liang et al. \cite{liang2017provchain}. For wireless provenance tracing, Kamal and Tariq developed a lightweight multi-hop system based on RSS indications \cite{kamal2018light}. Scalability, interoperability, and trust issues still arise when managing provenance across many platforms and sub-metaverses.

\subsection{Countermeasures to Technology Threats}
Metaverse experiences heavily rely on XR hardware (e.g., VR headsets, AR glasses, haptic devices), application interfaces, and rendering engines. All these devices and technologies can introduce distinct attack surfaces. Vulnerabilities in firmware, device drivers, or companion mobile apps can enable attackers to compromise user devices, exfiltrate data, or manipulate immersive environments. We present existing countermeasures to protect devices, applications, and user interfaces from attacks. 

\subsubsection{Secure Metaverse Devices and Games} 
AR/VR games collect user data to provide users with immersive experiences, but the collection of such personal information poses several security threats. According to Bono et al. \cite{bono2009reducing}, user devices can be compromised by exploiting MMO games such as Second Life and Minecraft. Lebeck et al. \cite{lebeck2018towards} highlight how immersive AR causes users to treat virtual objects as real, raising privacy risks. Shang et al. \cite{shang2020arspy} demonstrate a location tracking attack in AR games using network traffic, validated through real-user experiments. To reduce risks, Corcoran and Costache \cite{corcoran2018privacy} differentiate between individual and group privacy problems in game design, while Laakkonen et al. \cite{laakkonen2016incorporating} provide privacy-by-design principles. 

\subsubsection{Secure Device Firmware Updates}
Secure device firmware updates are a critical defense mechanism in the metaverse ecosystem. Updated device firmware ensures that VR/AR headsets, haptic devices, and other connected hardware operate with the latest security patches \cite{ruth2019secure}. Outdated firmware can be exploited by attackers to insert malicious code, manipulate device behavior, or steal sensitive data. Implementing cryptographically signed updates prevents unauthorized modifications and ensures device integrity \cite{meta-privacy-survey:wang2022survey}. Automated update pipelines, coupled with rollback protection, can further reduce downtime and prevent adversaries from exploiting vulnerabilities in legacy versions of virtual systems.

\subsubsection{Sandboxing XR Applications}
Sandboxing XR programs limits their access to user data and sensitive system resources by isolating them in contexts with controlled execution \cite{tennakoon2022teaching}. As third-party XR apps are crucial to the metaverse's content generation, engagement, and commerce, sandboxing is necessary to prevent dangerous activity by compromised or unconfirmed programs. 
By enforcing strict permission models and runtime monitoring, sandboxing can prevent unauthorized file system access, keystroke logging, or data harvesting within virtual environments \cite{falchuk2018social, tennakoon2022teaching}. This method not only reduces the impact of an attack but also makes post-breach forensic analysis easier.

\subsubsection{API Security Best Practices}
In Metaverse systems, APIs are the foundation for interoperability, facilitating communication between devices and applications \cite{oh2023secure}.  On the other hand, insecure APIs can be used to interrupt services, steal accounts, or compromise data \cite{han2010user}. Following proper API development guidelines, strong authentication, access control, input validation, rate limiting, and comprehensive logging can be significantly reduced.  Adopting a security standard such as OAuth 2.0  can stop unwanted access to user data and services, while using mutual TLS for API communication guarantees confidentiality and integrity \cite{han2010user}.

\subsubsection{Zero-Trust Device Policies}
No device, whether internal or external to the metaverse network, should be assumed to be trustworthy by default, according to zero-trust device policies \cite{meta-privacy-survey:wang2022survey}. All access requests are constantly permitted, encrypted, and authenticated, irrespective of the device's location or prior trust status \cite{he2022survey}. Zero-trust rules ensure that compromised XR headsets, IoT sensors, or user endpoints can not move laterally within the network to escalate privileges or steal data in the context of metaverse installations. Maintaining system integrity requires stringent enforcement of behavioral analytics, posture evaluations, and device health checks \cite{he2022survey}.

\subsubsection{Network Threats}
Protecting Metaverse platforms from DDoS attacks, wormhole attacks, and other network-level threats is crucial at the infrastructure level. Kuo (2023) introduced WD-SPRT, a lightweight wormhole detection system based on statistical anomaly detection. It uses fluctuations in network neighbor counts to flag suspicious routing patterns in mobile cloud-based Metaverse setups. This approach is perfect for scalable, real-time detection and doesn't require any more hardware. According to Bhardwaj et al. \cite{bhardwaj2023metaverse}, systems based on generative adversarial networks (GANs) and long short-term memory (LSTM) neural networks are being developed for broader intrusion prevention in order to monitor traffic patterns and identify anomalies in decentralized systems.
Using XR technology to improve local awareness is the topic of recent studies \cite{woodward2022analytic, ju2020acoustic, lv2020industrial}. Woodward and Ruiz \cite{woodward2022analytic} investigate how AR can enhance user perception. 
According to Ju et al. \cite{ju2020acoustic}, audio signals in virtual reality enhance emergency reaction in driving simulations. To detect unknown attacks, Vu et al. \cite{vu2020learning} propose using regularized autoencoders to learn latent attack representations. Zhang et al. \cite{zhang2013medmon} offer a lightweight anomaly detection method leveraging radio-frequency communications in wearables. 
In order to manage overfitting with little labeled data, Zhou et al. \cite{zhou2020siamese} combine few-shot learning with Siamese networks, while \cite{heartfield2020self} investigates RL-based intrusion detection for smart homes.

\subsection{Countermeasures Governance Threats}
The governance of the Metaverse faces critical vulnerabilities such as regulatory misconduct, centralization of control, jurisdiction ambiguity, and regulator collusion. Several countermeasures have been proposed to address these challenges. 
For large-scale systems, decentralized governance models (both flat and hierarchical) are preferred in order to improve resilience and avoid the concentration of regulatory power \cite{goldberg2023metaverse, almeida2021ecosystem}. Collaborative frameworks involving multiple stakeholders can help distribute decision-making authority and improve transparency. 
However, to prevent collusion among regulators, trust evaluation processes and anomaly detection systems are essential \cite{goldberg2023metaverse}. Smart contracts provide a technical means to automate governance procedures and enforce rules, reducing human intervention and guaranteeing compliance \cite{rahaman2024meta}. Verifiable and effective dispute resolution across jurisdictional borders can be achieved using blockchain-based arbitration systems \cite{al2025navigating}. Additionally, distributed validator committees and modular governance structures are suggested as solutions to the scalability and adaptability requirements of metaverse systems \cite{goldberg2023metaverse}.  Collectively, these countermeasures aim to build a robust and fair governance ecosystem for the Metaverse.

\begin{table*}[htbp]
\centering
\tiny
\caption{Countermeasures for Security Vulnerabilities in the Metaverse (RQ4)}
\label{tab:rq4-summary}
\renewcommand{\arraystretch}{1.3}

\begin{tabularx}{\linewidth}{@{}p{2.3cm}p{2.8cm}X p{2.5cm}@{}}
\toprule
\textbf{Category} & \textbf{Sub-category} & \textbf{Countermeasures} & \textbf{Validation Level} \\
\midrule

\multirow{6}{*}{Human}

& Identity and Authentication 
& Blockchain-based access control \cite{yao2022metaverse, seo2024space}, biometric authentication \cite{zhang2024anti}, cross-platform MFA \cite{shen2020blockchain, chen2021xauth} 
& Validated \\

& Access Control 
& Smart contracts and multi-signature authentication \cite{gai2022blockchain} 
& Validated \\

& Privacy and Data Protection 
& Privacy-preserving UGC processing \cite{wei2020ldp, zhang2021fedsens, guan2020blockmaze, wang2020spds}, encryption at rest and in transit \cite{oh2023secure} 
& Validated \\

& Identity Protection 
& Avatar cloaking techniques \cite{yang2023secure, cheong2022avatars} 
& Prototyped \\

& Content Security 
& AI-driven moderation \cite{ruth2019secure, dickinson2021experiencing}, deepfake detection \cite{tariq2023deepfake}, misinformation filtering \cite{zhu2020activity, sun2020stealthy} 
& Validated / Prototyped \\

& Secure Data Sharing 
& Secure data sharing frameworks \cite{ruth2019secure, lee2021adcube} 
& Validated \\

\midrule

\multirow{6}{*}{Technology}

& XR Security 
& Secure firmware updates \cite{ruth2019secure}, sandboxing XR apps \cite{tennakoon2022teaching} 
& Validated / Prototyped \\

& Network Security 
& Anomaly detection in 5G/6G \cite{zhang2013medmon, vu2020learning}, DDoS mitigation \cite{yan2018multi}, secure network slicing \cite{de2023survey} 
& Validated \\

& Digital Twin and System Security 
& API security \cite{oh2023secure}, secure synchronization \cite{liao2021digital}, zero-trust architectures \cite{he2022survey} 
& Prototyped \\

& AI and Data Security 
& Adversarial robustness \cite{sun2020stealthy}, data poisoning defense \cite{otoum2024machine}, federated learning \cite{li2021verifiable} 
& Validated / Prototyped \\

& Blockchain and Smart Contract Security
& Formal verification \cite{sayeed2020smart}, vulnerability scanning \cite{rahaman2024meta}, secure consensus \cite{alkhalifah2021mechanism}  
& Validated / Prototyped \\

& IoT and Device Security 
& Device attestation \cite{kim2022novel}, secure boot \cite{han2022dynamic}, IoT IDS \cite{shen2020blockchain}, RL-based defense \cite{heartfield2020self} 
& Validated \\

\midrule

\multirow{2}{*}{Governance \& Social}

& Governance Frameworks
& Transparent governance \cite{almeida2021ecosystem}, smart contract-based regulation \cite{rahaman2024meta}, accountability mechanisms 
& Theoretical / Prototyped \\

& Trust and Accountability 
& Decentralized identity \cite{goldberg2023metaverse}, cross-platform trust models \cite{almeida2021ecosystem}, audit trails \cite{almeida2021ecosystem}
& Theoretical \\

\bottomrule
\end{tabularx}
\end{table*}

\begin{tcolorbox}[boxrule=0.5pt,boxsep=1pt,left=2pt,right=2pt,top=2pt,bottom=2pt]
\noindent\textbf{RQ3 Summary:}
Securing the Metaverse requires a multi-layered, interdisciplinary approach that addresses identity verification, privacy, data integrity, behavioral safety, and system-level robustness. Blockchain-based identity frameworks, multi-factor biometric authentication, AI-driven anomaly detection,  deepfake detection algorithms, misinformation filtering, and avatar privacy tools are examples of defense mechanisms that are presently being developed or implemented. 
Table \ref{tab:rq4-summary} summarizes the existing SOTA countermeasures for security threats and vulnerabilities in Metaverse. 
\end{tcolorbox}

\section{Research Challenges and Future Directions for Metaverse (RQ4)}  \label{sec:rq5}
The rise of the Metaverse represents a revolutionary shift in how people interact digitally. Metaverse combines immersive virtual environments with real-world data, decentralized economies, and intelligent agents to provide users with a hyper-realistic environment. Although trust and security are essential for its long-term adoption, maintaining them in such a complex, multidimensional ecosystem is still a difficult undertaking. Decentralized governance, AI-based threat detection, identity management, and authentication protocols have all advanced, yet several research challenges still exist. Challenges such as technical, ethical, legal, and socio-economic domains require interdisciplinary investigation to establish secure and trustworthy Metaverse environments.

\mypara{1. Standardization and Interoperability of Security Protocols:}
The absence of common security standards across many platforms, XR devices, and blockchain infrastructures is a major barrier to Metaverse security. Access control, content moderation, and identity verification are implemented differently across platforms, resulting in fragmentation and inconsistency. This lack of interoperability expands the attack surface and makes it difficult to implement unified defenses against attacks. Open research questions include: \textit{How can decentralized identity frameworks (such as DIDs and verifiable credentials) be made interoperable across platforms?} \textit{Can blockchain-based authentication protocols (e.g., Metaverse-AKA) be standardized to support cross-platform access without compromising privacy?} Developing interoperable security layers and establishing international standards is very important, but remains underexplored. 

In order to ensure compatibility across virtual environments, interoperable security frameworks built on open standards could be utilized. Adoption of API standardization and cross-platform identity federation can facilitate easy authorization and authentication. Furthermore, blockchain-based interoperability layers can enable safe asset and identity transfer across platforms, while preserving consistency and trust.

\mypara{2. Secure and Privacy-Preserving Biometric Authentication:}
Continuous collection of extremely sensitive biometric data, including voice, body posture, facial expressions, iris scans, and behavioral patterns, is part of the Metaverse's use of XR devices. Authentication, personalization, and behavioral modeling are common uses for these data, but they carry serious privacy hazards. Although there have been significant advancements, such as the combination of facial and body biometrics for strong avatar verification \cite{zhang2024anti}, there are still issues with the current systems: \textit{Once compromised, biometric data cannot be undone and cannot be recovered easily}; \textit{Deepfake content spoofing attacks can still get past a lot of modern systems}; \textit{ Consent-driven, selective data sharing is not well supported on XR platforms}. Developing biometric authentication methods that are safe, reversible, and privacy-preserving and that can be effectively implemented on XR devices must be the main focus of future research.

On-device biometric processing can reduce privacy threats by guaranteeing that raw data never leaves the device of the user. Biometric templates can be further protected using methods such as homomorphic encryption \cite{ogburn2013homomorphic} and differential privacy \cite{dwork2025differential}. Furthermore, security can be improved by using behavioral biometric-based authentication.

\mypara{3. Input Provenance and Trustworthy Interaction Tracking:}
Distinguishing authentic user input from synthetic or adversarial input is a significant challenge in the 3D, multi-sensory interfaces of the Metaverse. Clickjacking, object deletion, and synthetic input injection are examples of UI-level attacks that might make users unintentionally perform harmful actions. Cheng et al. (2024) highlighted that there is a basic deficiency in "input provenance," which is a trustworthy method of determining the origin of a particular interaction and its authenticity \cite{cheng2024user}. Open challenges include: \textit{Developing real-time provenance verification systems for XR interfaces}; \textit{Detecting and blocking adversarial inputs without impairing user experience}; and \textit{Creating standardized logging mechanisms for forensics and accountability in immersive environments}.

Data integrity and origin authenticity can be guaranteed by cryptographic techniques such as digital signatures \cite{gai2022blockchain} and hash-based verification \cite{wang2017verification}. Traceability can be achieved by secure time-stamping and audit trails, while blockchain-based logging systems can offer recordings of interactions. These can be combined with trusted hardware to guarantee secure generation and verification of input data.

\mypara{4. Content Moderation in Decentralized Environments:}
Since Metaverse sites are decentralized and frequently run on blockchain, centralized content moderation is either impractical or ineffectual. However, these platforms are vulnerable to deepfakes, hate speech, harassment, and disinformation abuse. The difficulty lies in developing real-time, context-aware, scalable moderation systems that work in user-generated, immersive, decentralized environments. Important questions are: \textit{Can AI-powered moderation systems identify harmful behavior in real-time without bias?} \textit{How can moderation policies be transparently and fairly implemented in decentralized governance models?} \textit{ What ethical frameworks should guide automated moderation in immersive spaces?} Collaboration between the communities of computer science, law, and ethics is necessary for these active research areas.

Scalability and transparency can be increased with a hybrid strategy that combines community-based governance frameworks with AI-driven moderation. Furthermore, human-in-the-loop validation for crucial judgments supports that automated filtering techniques can be used for real-time detection of dangerous information.

\mypara{5. Defense Against Deepfake and AI-Generated Threats: }
In the Metaverse, deepfakes and generative AI tools are becoming more dangerous, especially in social, academic, and professional settings. Attackers can use incredibly lifelike synthetic media to impersonate, manipulate digital identities, or propagate false information. Although watermarking and GAN-detection models have been used as detection methods, they are not highly accurate and reliable. Future studies should address: \textit{How to detect deepfakes in real time during immersive interactions?} \textit{How to watermark or authenticate avatars and media?} \textit{What legal safeguards should exist against deepfake misuse in professional and governmental contexts?} This research area also involves combating adversarial machine learning attacks, where models are tricked into misclassification.

Deepfake detection algorithms, content watermarking, and media provenance systems are examples of mitigation techniques. Traceability can be ensured by incorporating cryptographic signatures into the created content. Additionally, real-time verification pipelines can be used to stop fraudulent content from spreading.

\mypara{6. Secure Federated Learning and Collaborative AI:}
AI models in the Metaverse rely on vast amounts of user data for training and inference. Federated learning (FL), in which training takes place privately on user devices and only model updates are shared, is being used more and more to safeguard user privacy. However, hostile participants and poisoning assaults can affect FL itself. Future directions include \textit{developing robust federated learning protocols resilient to data poisoning; }\textit{ensuring the integrity of collaborative AI without requiring centralized control}; \textit{Investigating adversarial attacks on model updates and proposing defense mechanisms.} Li et al. studied adversary deactivation in FL, highlighting the potential, but broader, scalable solutions are still needed \cite{li2023robust}.

Robust aggregation approaches (e.g, Byzantine-resilient aggregation) can reduce harmful updates and secure these systems. During training, data confidentiality can be protected using differential privacy. Furthermore, trust score systems can control contributions in collaborative settings, and model validation and anomaly detection techniques can uncover compromised systems.

\mypara{7. Avatar Integrity and Social Engineering Defense:}
In the Metaverse, avatars serve as social identities as well as functional proxies. Attackers can carry out identity theft, harassment, or social engineering attacks by cloning avatars, imitating behavior, or manipulating interactions. Open research areas include \textit{developing secure avatar creation and verification protocols}; \textit{preventing behavioral fingerprinting through motion data obfuscation}; and \textit{creating user-friendly tools for blocking, reporting, and recovering from avatar hijacking}. Moreover, the moral and legal implications of avatar impersonation and behavioral inference remain underdefined.

Cryptographic identity binding can be used to verify user identities. Behavioral biometrics (e.g., movement patterns and interaction styles) can be combined with the multi-factor and continuous authentication methods that can potentially help to identify impersonation attempts.

\mypara{8. Darkverse and Untraceable Threats:}
Darkverse is a private, encrypted, decentralized virtual environment and is emerging as a major concern. These environments can serve as havens for criminal activity, uncontrolled trade, and radicalization outside the jurisdiction of law enforcement or moderation. Research challenges include: \textit{Identifying indicators of misuse in encrypted, pseudonymous environments; Developing surveillance-resistant yet accountable architectures}; and \textit{Balancing privacy with legal obligations and societal safety}. Cybersecurity, encryption, and policy must work together in this delicate area to address legitimate concerns without violating rights.

To solve darkverse and untraceability, privacy-conscious traceability techniques such as zero-knowledge proofs and selective disclosure credentials are required. Without depending only on identity, behavioral anomaly detection systems can be utilized to identify unusual activity.

\mypara{9. Legal and Ethical Frameworks for Trust:}
User-centered, transparent, and enforceable legal frameworks are necessary to ensure trust in the Metaverse. Principal issues include: \textit{who is legally responsible for actions taken by avatars?} \textit{How should digital assets and identities be protected under law?} and \textit{What rights do users have to privacy, safety, and redress in virtual environments?} Researchers and legal scholars must work together to establish ethical design standards, culpability, ownership, and digital personality.

To ensure uniform enforcement, common regulatory frameworks should be developed across countries. While auditable governance structures enhance transparency, smart contract-based compliance methods can automate policy enforcement. Furthermore, ethics-by-design, or integrating moral principles into system design, promotes ethical technology deployment.

\mypara{10. Sustainability and Energy-Aware Security:}
The Metaverse's security systems need a lot of energy, particularly those that use edge rendering, AI processing, and real-time encryption. The development of resource-efficient, lightweight protocols that work on mobile and XR devices is necessary to ensure sustainable security. Open questions are: \textit{Is it possible to make energy-efficient secure authentication and encryption for wearables? }\textit{How can decentralized networks strike a balance between power usage, security, and performance?} \textit{How much does a secure Metaverse infrastructure cost the environment, and how can that cost be reduced?} Sustainable trust architectures and green security are two understudied but increasingly important topics.

Lightweight cryptographic algorithms, energy-efficient consensus techniques, and edge-based processing to minimize data transit can increase energy efficiency. In order to maintain a balance between security and sustainability, adaptive security measures can also dynamically modify resource utilization based on threat levels.











\begin{tcolorbox}[boxrule=0.5pt,boxsep=1pt,left=2pt,right=2pt,top=2pt,bottom=2pt]
\noindent\textbf{RQ4 Summary:}
The development of secure and trustworthy Metaverse environments presents a multiplicity of open research challenges that span technical, ethical, and legal limits. Avatar integrity, federated AI security, privacy-preserving biometrics, deepfake detection, protocol standardization, decentralized moderation, and sustainability represent key challenges. Researchers, businesses, policymakers, and civil society must work together to address these issues. 

\end{tcolorbox}
\section{Limitations} \label{sec:limitations}
This survey provides a thorough overview of technologies, security, privacy, and trust issues in the Metaverse. However, there are several limitations that remain.
First, given how quickly metaverse technologies are developing, some new dangers and defenses might not be completely covered, particularly if they are made available after the publication date.
Second, despite our efforts to provide comprehensive coverage, there will inevitably be gaps in platform-specific information and proprietary solutions due to the variety of platforms, architectures, and implementation methods. 
Third, certain countermeasures are not as practically applicable as the majority of current research focuses on theoretical models or simulations with little validation in the actual Metaverse environment.
Fourth, due to the abundance of research on technical aspects, the focus has shifted more in that direction, whereas less emphasis has been placed on the legal, ethical, and socio-cultural dimensions of Metaverse security and privacy. Finally, the lack of standardized metrics and benchmarking tools across studies makes it difficult to compare the effectiveness of proposed solutions consistently. These restrictions provide new avenues for investigation and security framework improvement in the dynamic Metaverse environment.

\section{Conclusion} \label{sec:conclusion}
The Metaverse is transforming the digital paradigm by combining physical and virtual worlds. It uses advanced technologies such as XR, blockchain, AI, and digital twins to provide a realistic environment for the users. In this survey, we presented a comprehensive overview of the foundational technologies enabling the Metaverse. We also analyze the associated security, privacy, and trust issues that arise in Metaverse environments.  We have analyzed a wide range of vulnerabilities and their corresponding countermeasures. In addition, we discussed countermeasures that utilize authentication frameworks, cryptographic methods, decentralized architectures, and AI-driven detection systems to ensure robust and resilient Metaverse platforms. Many suggested solutions are still in the conceptual or prototype stage, with little implementation in real-world settings, despite the encouraging advancements. In addition, challenges such as scalable data provenance, ethical governance, cross-domain interoperability, and demands for individualized privacy remain major obstacles. Strong cooperation from academics, business, and policymakers is needed to overcome these unresolved issues as the Metaverse ecosystem grows. Future research should focus on integrating multi-disciplinary approaches, fostering standardization, and validating proposed frameworks in real-world applications.



%


\section*{Acknowledgment}
This research is supported by the National Cybersecurity Consortium and the Canadian Institute for Cybersecurity at the University of New Brunswick.

\ifCLASSOPTIONcaptionsoff
  \newpage
\fi



%

\bibliographystyle{unsrt}
\bibliography{ref}


%










\end{document}